\documentclass[fleqn,useAMS,usenatbib]{mnras}
\usepackage{myaasmacros}
\usepackage{booktabs,tabularx,graphicx,caption,amsmath,wasysym,makecell,hhline,enumitem,fancyvrb,gensymb,float}
\usepackage{hyperref} % <-- I had to remove this due to bugs!
\hypersetup{draft}
\usepackage{ulem, color, bm}

\def\ltsima{$\; \buildrel < \over \sim \;$}
\def\simlt{\lower.5ex\hbox{\ltsima}}   
\def\gtsima{$\; \buildrel > \over \sim \;$}
\def\simgt{\lower.5ex\hbox{\gtsima}}

\title[Globular clusters as probes of dark matter cusp-core transformations]{Globular clusters as probes of dark matter cusp-core transformations}

\author[M. D. A. Orkney; J. I. Read; James A. Petts; Mark Gieles]{M. D. A. Orkney$^{1}$\thanks{E-mail: m.orkney@tiscali.co.uk}, J. I. Read$^1$, J. A. Petts$^1$, M. Gieles$^{1,2,3}$\\
$^1${\small Department of Physics, University of Surrey, Guildford, GU2 7XH, Surrey, UK}\\
$^2$Institut de Ci\`{e}ncies del Cosmos (ICCUB), Universitat de Barcelona, Mart\'{i} i Franqu\`{e}s 1, 08028 Barcelona, Spain\\
$^3$ICREA, Pg. Lluis Companys 23, 08010 Barcelona, Spain.
}

\begin{document}

\maketitle

% Things to do for the proof:
% Make sure that I fix the references in the introduction as per the advice given to me over Twitter. \citep{2017ApJ...837...54C} (many authors). McGaugh stuff: \citep{2005ApJ...621..757S} (many authors), \citep{2008AJ....136.2648D} (many authors), \citep*{2008ApJ...676..920K} (3 authors!)

% Make sure that I include a reference to the 6th new GC in Fornax. Don't include this in Figures or analysis. \citep{2019ApJ...875L..13W} (many authors) <-- I stuffed this in the first footnote. Appropriate or not? Hmm.

% Make sure that I add reference for a new large GC in a dwarf galaxy! \citep{2019MNRAS.487.1986B} (many authors) <-- Stuck at the end of the "GCs evolve to larger sizes in a cored background potential" section.

% Make sure I add this paper: \citep{2019MNRAS.485.2546B} which suggests the GCs in Fornax are compatible with an NFW profile. I also need to reply to their email. <-- Stuffed this in the discussion section, but I should also read their paper to ensure it is sensible!

\begin{abstract}
Bursty star formation in dwarf galaxies can slowly transform a steep dark matter cusp into a constant density core. We explore the possibility that globular clusters (GCs) retain a dynamical memory of this transformation. To test this, we use the {\sc nbody6df} code to simulate the dynamical evolution of GCs, including stellar evolution, orbiting in static and time-varying potentials for a Hubble time. We find that GCs orbiting within a cored dark matter halo, or within a halo that has undergone a cusp-core transformation, grow to a size that is substantially larger ($R_{\rm eff} > 10$\,pc) than those in a static cusped dark matter halo. They also produce much less tidal debris. We find that the cleanest signal of an historic cusp-core transformation is the presence of large GCs with tidal debris. However, the effect is small and will be challenging to observe in real galaxies. Finally, we qualitatively compare our simulated GCs with the observed GC populations in the Fornax, NGC 6822, IKN and Sagittarius dwarf galaxies. We find that the GCs in these dwarf galaxies are systematically larger ($\langle R_{\rm eff}\rangle \simeq 7.8$\,pc), and have substantially more scatter in their sizes, than in-situ metal rich GCs in the Milky Way and young massive star clusters forming in M83 ($\langle R_{\rm eff} \rangle \simeq 2.5$\,pc). We show that the size, scatter and survival of GCs in dwarf galaxies are all consistent with them having evolved in a constant density core, or a potential that has undergone a cusp-core transformation, but not in a dark matter cusp.

\end{abstract}


\section{Introduction} \label{sec:intro}

Our standard cosmological model, $\Lambda$ Cold Dark Matter ($\Lambda$CDM), accurately describes the cosmic microwave background radiation \citep{2009ApJS..180..306D, 2017JCAP...04..012S}, galaxy clustering \citep{1998Natur.392..359G} and lensing of galaxy clusters (\citealp{1991MNRAS.251..600B, 1993ApJ...404..441K}; \citealp*{2002ApJ...574L.129S}). Pure dark matter $N$-body simulations of structure formation in $\Lambda$CDM suggest that dark matter (DM) halos have a self-similar `universal' density distribution, described by the `NFW' profile \citep*{1996ApJ...462..563N}:
\noindent\begin{equation} \label{NFW.eq}
\rho_{\rm NFW}(r) = \frac{\rho_{0}}{r/r_{\rm s}\left ( 1+r/r_{\rm s} \right )^{2}},
\end{equation}
where $\rho_{0}$ is a density normalisation, $r$ is the radius from the galaxy centre, and $r_{s}$ is the scale radius. For small radii, $r\ll r_s$, $\rho_{\rm NFW} \simeq\rho_0 ({r}/{r_s})^{-1}$ which is called a DM `cusp' since the density diverges as $r \rightarrow 0$. However, there has been a long-standing tension between the above prediction of a divergent cuspy density profile and observations of the rotation curves of nearby isolated dwarf irregular galaxies \citep{1994ApJ...427L...1F,1994Natur.370..629M,2017MNRAS.467.2019R}. These favour instead an inner region where the density is constant, $\rho \propto r^{0}$. This disparity is known as the `cusp-core problem'. 

One solution to the cusp-core problem is that baryonic processes, missing in the \citet{1996ApJ...462..563N} simulations, are somehow responsible for coring an originally cuspy DM halo, an effect that has become known as `dark matter heating'. Candidate mechanisms include dynamical friction (\citealp*{2001ApJ...560..636E}; \citealp{2008MNRAS.386.2194N, 2009A&A...502..733D, 2015MNRAS.446.1820N}) and stellar feedback from recurrent starbursts \citep{1996ApJ...462..563N, 2005MNRAS.356..107R, 2012MNRAS.421.3464P, 2014Natur.506..171P} to which dwarf galaxies are particularly susceptible due to their shallow potential wells. Indeed, we now have observational evidence for this dark matter heating operating in dwarf galaxies. \citet{2019MNRAS.484.1401R} find an anti-correlation between the central dark matter density of a sample of 16 nearby dwarfs and their total star formation, consistent with predictions from recent dark matter heating models (\citealp[e.g.][]{2012ApJ...759L..42P,2014MNRAS.437..415D,2015MNRAS.454.2981C}; \citealp*{2016MNRAS.459.2573R}). \par

In this paper, we ask whether we can find evidence for dark matter heating transforming a dark matter cusp to a core in an {\it individual dwarf galaxy}. The key idea we explore is that the maximum radius globular clusters can expand to due to two-body relaxation depends on the tidal field they inhabit (\citealp{1961AnAp...24..369H}; \citealp*{2011MNRAS.413.2509G}; \citealp{2018MNRAS.476.3124C,2018MNRAS.479.3708W}). \citet{2018MNRAS.476.3124C} have recently used this idea to argue for the presence of a dark matter core in the nearby ultra-faint dwarf galaxy Eridanus II. This is distinct from earlier works that have used the {\it survival} of globular clusters (GCs) to probe the presence or absence of a dark matter cusp \citep[e.g.][]{2006MNRAS.368.1073G,2010ApJ...725.1707G,2016MNRAS.461.4335A, 2017ApJ...844...64A}. In this paper, we take this idea further to explore whether the current properties of GCs could encode information about {\it historic} cusp-core transformations. To address this question, we simulate idealised GCs orbiting a galaxy with static and time-varying galactic potentials. We consider GCs with a range of initial sizes and masses given by $R_{\rm eff} \simeq$ 0.5, 1, 2\,pc and $M \simeq$ 20k, 40k, 80k\,M$_{\odot}$ respectively \footnote{See \citet[Fig. 2]{2007ApJS..171..101J} for the typical mass distribution of Galactic GCs, and \citet[Table 3]{2016A&A...590A..35D} for mass estimates of the GC system in the Fornax dSph. Note that recently, \citep{2019ApJ...875L..13W} have highlighted the likely presence of a 6\textsuperscript{th} GC in Fornax. The Fornax GCs have a similar range in mass ($0.42 \times 10^5 - 4.98 \times 10^5$\,M$_{\odot}$) to those we simulate here.}, and a range of static host galaxy potentials with different central logarithmic slopes corresponding to a core, a cusp, and something in-between. We also consider the case of a time-varying galactic potential that undergoes a cusp-core transformation. We base this time-varying potential on the simulations of isolated dwarfs from \citet{2016MNRAS.459.2573R} (hereafter R16), in which bursty star formation slowly transforms a DM cusp into a core. We use their simulation of an $M_{200} = 10^9$\,M$_\odot$ dwarf in which the cusp-core transformation time was $t_{\rm tr} = 8\,\text{Gyrs}$. In all cases, we consider planar circular orbits. We will consider the effect of elliptical orbits and/or non-spherical potentials in future work.

This paper is organised as follows. In \S\ref{sec:sims}, we discuss our simulation suite and we describe how we model the time-varying potential and the dynamical friction. In \S\ref{sec:results}, we present our results. In \S\ref{sec:discussion}, we discuss our results and compare them with recent work in the literature. We discuss the caveats inherent in our work and we qualitatively compare our findings to observations of GCs in nearby dwarf galaxies. Finally, in \S\ref{sec:conclusion}, we present our conclusions. \par

\section{The $N$-body simulations} \label{sec:sims}

The simulations were run using a variant of the {\sc nbody6} code \citep{1999PASP..111.1333A}, a Graphics Processing Unit (GPU) enabled \citep{2012MNRAS.424..545N} direct $N$-body simulation tool. This variant, named {\sc nbody6df} \citep*{2015MNRAS.454.3778P}, incorporates the effects of dynamical friction on GCs orbiting a host galaxy and is described in detail in \S\ref{df}. {\sc nbody6df} was altered as described in \S\ref{interpolation} to allow for a galactic cusp-core transformation which reproduces the time-varying gravitational potential, $\Phi(t)$, from R16. \par 

{\sc nbody6} uses regularisation to model binary, triple and higher-order stellar encounters in the GC. We did not include primordial binaries. Stellar evolution was modelled based on the `Eggleton, Tout and Hurley' option in {\sc nbody6} (\citealp*{1989ApJ...347..998E}; \citealp{1997MNRAS.291..732T}; \citealp*{2000MNRAS.315..543H, 2002MNRAS.329..897H}; \citealp{2008LNP...760..283H}). \par

\subsection{Initial conditions}

\begin{figure}
\setlength\tabcolsep{2pt}%
\includegraphics[ trim={0 1cm 0 0}, width=\columnwidth, height=\columnwidth, keepaspectratio]{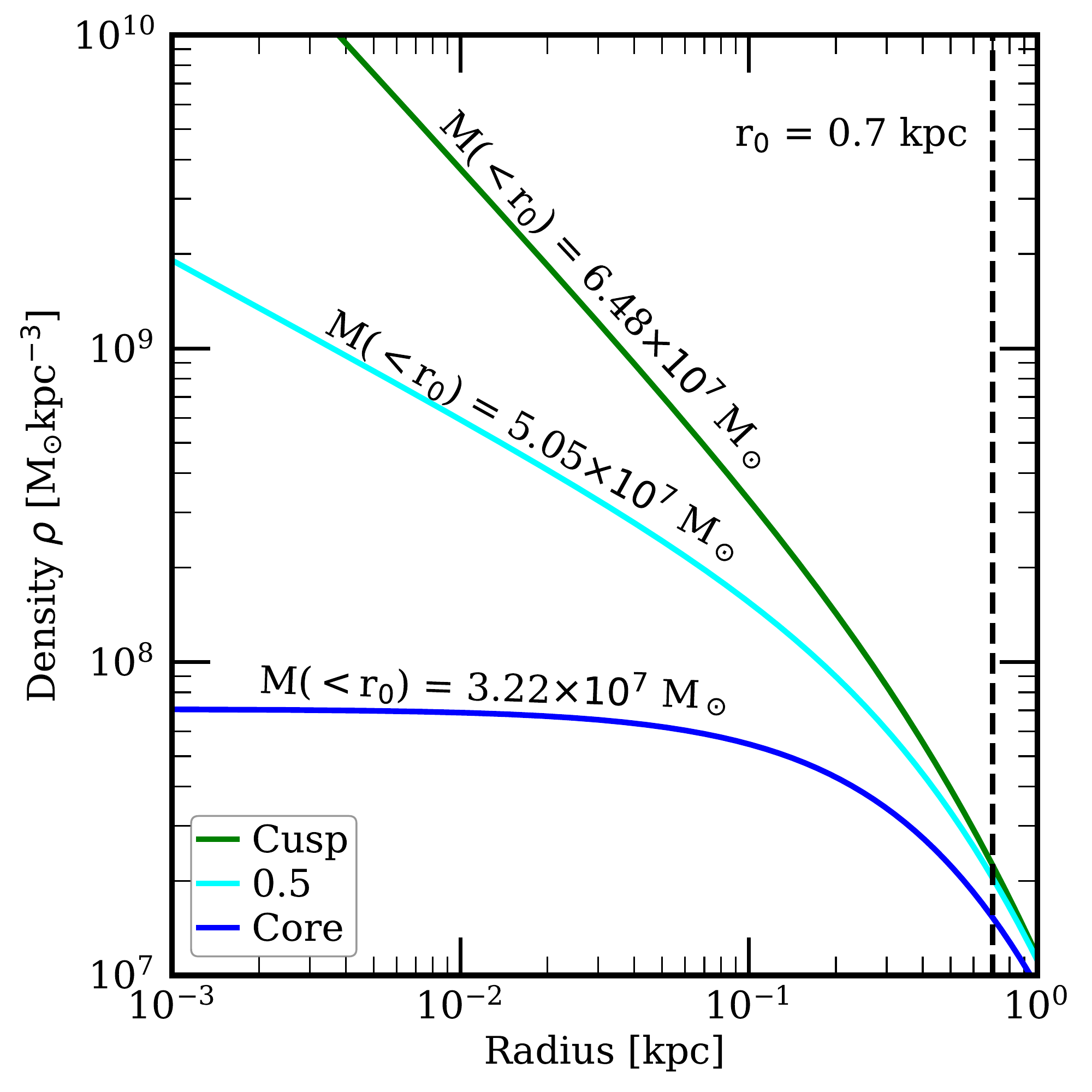}\\
\caption{Dehnen density profiles used to model the host galaxy potential. The dashed vertical line indicates the starting orbital radii of the GCs, $r_{0}=0.7$\,kpc, which was identical for each simulation. Each line is labelled with the mass enclosed at $r_{0}$.}
\label{fig:profiles}
\end{figure}

Our suite of simulations were designed to model the evolution of a GC orbiting a typical dwarf spheroidal (dSph) galaxy like Fornax \citep[e.g.][]{2006MNRAS.368.1073G}. Specifically, the galaxy properties were set up to mimic that of the $M_{200} = 10^9$\,M$_\odot$ galaxy presented in R16. For this, we assumed a spherically symmetric Dehnen density profile \citep{1993MNRAS.265..250D}:
\noindent\begin{equation}
\label{dehnen.eq}
\rho_{\rm Dehnen}(r) = \rho_{0} \left ( \frac{r}{r_{\rm s}} \right )^{-\gamma}  \left (1+\frac{r}{r_{\rm s}} \right )^{\gamma-4},
\end{equation}
where $\rho_{0}$ is the central density, $r_{\rm s}$ is the scale radius and $\gamma$ is a variable used to set the logarithmic slope of the inner density profile ($\gamma=0$ corresponds to a core and $\gamma=1$ corresponds to a cusp). This Dehnen profile was then fit to the simulations from R16, including a fit for the time-varying potential as detailed in \S\ref{interpolation}. (Note that we use a Dehnen profile here rather than the {\tt coreNFW} profile from R16 because the dynamical friction model in {\sc nbody6df} requires the full distribution function of the dark matter halo (see \S\ref{df}). For the Dehnen profile with $\gamma=0$ or $\gamma=1$, this is fully analytic, making the calculation significantly more efficient \citep{2015MNRAS.454.3778P}. We discuss this further in \S\ref{interpolation}.)

The GC initial conditions were set up with the {\sc McLuster} tool \citep{2011MNRAS.417.2300K} using a Plummer density model \citep{1911MNRAS..71..460P}, a Kroupa initial mass function (IMF) \citep{2010MNRAS.401..105K} and assuming no primordial binaries. The retention fraction of black holes (BHs) depends on the escape velocity from the centre of the cluster at the time of supernovae and the natal kicks of BHs. Both are poorly understood and here we adopt a BH retention fraction of 100\%, applicable to dense and massive GCs and zero natal kicks (i.e. as in the fallback scenario \citep{2012ApJ...749...91F}). The GC properties and orbital parameters were chosen to reflect typical GCs such as those observed in the dSph galaxy Fornax \citep{2012MNRAS.426..601C}. The largest GC was modeled with $N = 2^{17}$ ($\sim$128k) particles, which approaches the size of common GCs but is not so great as to exceed practical limits on computation time. The orbital velocity required to set a circular orbit ($V_{\rm circ}$) in a Dehnen potential can be derived from equation \ref{dehnen.eq} \citep{1993MNRAS.265..250D} as:
\noindent\begin{equation} \label{vcirc.eq}
  V_{\rm circ}^{2}(r)=\frac{GM_{\rm g}r^{2-\gamma}}{(r+r_{\rm s})^{3-\gamma}},
\end{equation}
where $M_{\rm g}$ is the mass of the galaxy and $G$ is the gravitational constant. Since our goal here is to compare the properties of idealised GCs moving in different potentials, we consider only circular orbits. The initial orbital radius was set to 700\,pc, which is similar to the orbital radii of GCs in comparable systems such as Fornax \citep{2012MNRAS.426..601C}. We will consider the effect of elliptical orbits in future work.

The suite consists of four main simulations in which the potential profile of the host dSph galaxy have different values of $\gamma$, as summarised in the upper section of Table \ref{tab:sim1}. The first three of these simulations use a constant $\gamma$ representing a core ($\gamma=0$), a cusp ($\gamma=1$) and an intermediate value ($\gamma=0.5$). The fourth simulation utilizes a variable $\gamma(t)$ fit to the simulations in R16 to mimic a cusp-core transformation. Profiles representing $\gamma=0$, $\gamma=1$ and $\gamma=0.5$ are shown in Fig. \ref{fig:profiles}. The profiles were designed to have a similar density at the initial GC orbital radii, $r_0$, marked by the vertical dashed line. The method by which $\gamma(t)$ was calculated is explained in \S\ref{interpolation}. \par 

Two sets of subsidiary simulations were run using identical properties but lower particle numbers, $N = 2^{16}$ and $N = 2^{15}$. The initial GC density was different for these sets of simulations because the particle count was reduced whilst maintaining the same initial projected half stellar mass radius. We will refer to the projected half radius as $R_{\rm eff}^\mathcal{M}$ when mass-weighted and $R_{\rm eff}^\mathcal{L}$ when luminosity-weighted, whereas we refer to the 3D half-mass radius as $R_{1/2}$.
\par

\newcolumntype{Y}{>{\centering\arraybackslash}X}
\begin{table*}
\captionof{table}{Principal parameters for each simulation. From left to right are the simulation name, the number of simulation particles, the initial GC orbital radius, the host galaxy mass, the variable $\gamma$ as in equation \ref{dehnen.eq}, the GC scale radius, the half-mass radius, the initial GC density, the initial orbital velocity and the initial GC relation time. The naming convention comprises the approximate particle number, the dwarf potential profile scheme employed and the initial density choice when relevant. The half-mass radius is the projected (2D) $R_{\rm eff}^\mathcal{M}$, which best compares with observation. The initial density is given by the average density within $R_{1/2}$. Parameters are to 3.s.f. or 3.d.p. as appropriate.} 
\begin{tabular}{lccccccccc}
 \toprule
 Name & Particles $N$ & $r_{0}$ & $M_{\rm galaxy}$ & $\gamma$ & $r_{\rm s}$ & $R_{\rm eff}^\mathcal{M}$ & $\rho_{0}(<R_{1/2})$ & $V_{\rm circ}$ & $t_{\rm relax,0}$\\
  & & (kpc) & (M$_{\odot}$) & & (kpc) & (pc) & (M$_{\odot}/pc^{3}$) & (km/s) & (Myr) \\
 \midrule
 128kcusp     & $2^{17}$  & 0.7  & $10^{9}$  & 1.0  & 2.05  & 0.963 & 4530   & 19.955  & 132.619   \\ 
 128k0.5      & $2^{17}$  & 0.7  & $10^{9}$  & 0.5  & 1.61  & 0.963 & 4530   & 17.626  & 132.619   \\
 128kcore     & $2^{17}$  & 0.7  & $10^{9}$  & 0.0  & 1.50  & 0.963 & 4530   & 14.070  & 132.619   \\ 
 128kvarying  & $2^{17}$  & 0.7  & $10^{9}$  & $\gamma(t)$ & 2.05  & 0.963 & 4530   & 19.955  & 132.619 \medskip  \\
 64kcusp      & $2^{16}$  & 0.7  & $10^{9}$  & 1.0  & 2.05  & 0.930 & 2440   & 19.955  & 99.377   \\
 64k0.5       & $2^{16}$  & 0.7  & $10^{9}$  & 0.5  & 1.61  & 0.930 & 2440   & 17.626  & 99.377   \\
 64kcore      & $2^{16}$  & 0.7  & $10^{9}$  & 0.0  & 1.50  & 0.930 & 2440   & 14.070  & 99.377   \\
 64kvarying   & $2^{16}$  & 0.7  & $10^{9}$  & $\gamma(t)$ & 2.05  & 0.930 & 2440   & 19.955  & 99.377 \medskip  \\
 32kcusp      & $2^{15}$  & 0.7  & $10^{9}$  & 1.0  & 2.05  & 0.973 & 1100    & 19.955  & 81.805   \\
 32k0.5       & $2^{15}$  & 0.7  & $10^{9}$  & 0.5  & 1.61  & 0.973 & 1100    & 17.626  & 81.805   \\
 32kcore      & $2^{15}$  & 0.7  & $10^{9}$  & 0.0  & 1.50  & 0.973 & 1100    & 14.070  & 81.805   \\
 32kvarying   & $2^{15}$  & 0.7  & $10^{9}$  & $\gamma(t)$ & 2.05  & 0.973 & 1100    & 19.955  & 81.805   \\
 \midrule
 64kcusplow   & $2^{16}$  & 0.7  & $10^{9}$  & 1.0  & 2.05  & 2.066 & 122    & 19.955  & 323.803  \\
 64k0.5low    & $2^{16}$  & 0.7  & $10^{9}$  & 0.5  & 1.61  & 2.066 & 122    & 17.626  & 323.803  \\
 64kcorelow   & $2^{16}$  & 0.7  & $10^{9}$  & 0.0  & 1.50  & 2.066 & 122    & 14.070  & 323.803  \\
 64kvaryinglow& $2^{16}$  & 0.7  & $10^{9}$  & $\gamma(t)$ & 2.05  & 2.066 & 122    & 19.955  & 323.803 \medskip  \\
 64kcusphigh    & $2^{16}$  & 0.7  & $10^{9}$  & 1.0  & 2.05  & 0.444 & 33000   & 19.955  & 32.612   \\ 
 64k0.5high     & $2^{16}$  & 0.7  & $10^{9}$  & 0.5  & 1.61  & 0.444 & 33000   & 17.626  & 32.612   \\
 64kcorehigh    & $2^{16}$  & 0.7  & $10^{9}$  & 0.0  & 1.50  & 0.444 & 33000   & 14.070  & 32.612   \\
 64kvaryinghigh & $2^{16}$  & 0.7  & $10^{9}$  & $\gamma(t)$ & 2.05  & 0.444 & 33000   & 19.955  & 32.612   \\
 \bottomrule
\end{tabular}
\label{tab:sim1}
\end{table*}

\subsection{Initial GC density}

A secondary suite of simulations was designed to investigate the sensitivity of results to the initial GC density, summarised in the lower section of Table \ref{tab:sim1}. Four simulations were run with different host galaxy potentials: $\gamma=0$, $\gamma=1$, $\gamma=0.5$ and $\gamma(t)$. A set of higher density and lower density GCs were then designed by varying the initial $R_{1/2}$. These radii were chosen such that the initial central stellar density was higher or lower by an order of magnitude than our fiducial suite of simulations.

For reasons of computational cost, this investigation into GC densities was performed only on the $N = 2^{16}$ simulations. As we shall see in \S\ref{sec:hmr}, the results are qualitatively similar for the $N = 2^{17}$ and $N = 2^{16}$ simulations. \par

\subsection{{\sc nbody6df}} \label{df}

Dynamical friction is a drag force imparted on a body as it moves through a sea of lighter background bodies \citep{1943ApJ....97..255C}. GCs experience dynamical friction as they orbit through their host galaxy as a result of interactions with stars, interstellar gas and DM \citep[e.g.][]{1984MNRAS.209..729T, 1998MNRAS.297..517H, 2001ApJ...552..572L, bandt, 2016Ap&SS.361..162D}. In dense galaxies, dynamical friction continues until the GC is tidally destroyed or it reaches the galactic centre. However, in galaxies with a central constant density core, dynamical friction stalls inside the core region (\citealp{2006MNRAS.373.1451R, 2006MNRAS.368.1073G, Inoue09, 2011MNRAS.416.1181I, 2015MNRAS.454.3778P}; \citealp*{Petts16}; \citealp{2018ApJ...868..134K}). %

In this paper, we use the semi-analytic model for dynamical friction implemented in {\sc nbody6df} \footnote{{\sc nbody6df} is publicly available and can be downloaded from \href{https://github.com/JamesAPetts/NBODY6df}{https://github.com/JamesAPetts/NBODY6df}.}. The method is described in detail in \citet{2015MNRAS.454.3778P, Petts16}. Here, we briefly summarise the main points. The frictional deceleration is given by:
\noindent\begin{equation}
	\frac{{\rm d}{\bm v}_{\rm s}}{{\rm d}t} = -2\pi G^2 M_{\rm s} \rho \log(\Lambda^2 + 1) f(v_{*} < v_{\rm s}) \frac{{\bm v}_{\rm s}}{{v^3_{\rm s}}},
	\label{dynfric.eq}
\end{equation}
where ${\bm v}_{\rm s}$ is the satellite velocity ($v_{\rm s} \equiv |{\bm v}_{\rm s}|$), $M_{\rm s}$ is the satellite mass, $\rho$ is the local background density, $f(v_{*} < v_{\rm s})$ is the distribution function representing the fraction of stars moving slower than the satellite and $\log(\Lambda^2 + 1)$ is the Coulomb logarithm $\Lambda$ given by:
\noindent\begin{equation}
	\Lambda = \frac{b_{\rm max}}{b_{\rm min}} = \frac{\mathrm{min}(\rho(R_{\rm g})/|\nabla\rho(R_{\rm g})|,R_{\rm g})}{\mathrm{max}\left(R_{1/2}, GM_{\rm s}/v_{\rm s}^2\right)},
	\label{coulog.eq}
\end{equation}
where $R_{\rm g}$ and $R_{1/2}$ are the galactocentric distance and 3D half-mass radius of the satellite, respectively. The satellite mass is based on the mass within its instantaneous Roche volume, rather than just the bound stars. This is because unbound stars that remain in the vicinity of the satellite are found to contribute to the dynamical friction force \citep{2007MNRAS.375..604F, 2015MNRAS.454.3778P}.  \par

\subsection{Modelling the cusp-core transformation, $\gamma(t)$} \label{interpolation}

\begin{figure}
\setlength\tabcolsep{2pt}%
\includegraphics[ trim={0 1cm 0 0}, width=\columnwidth, height=\columnwidth, keepaspectratio]{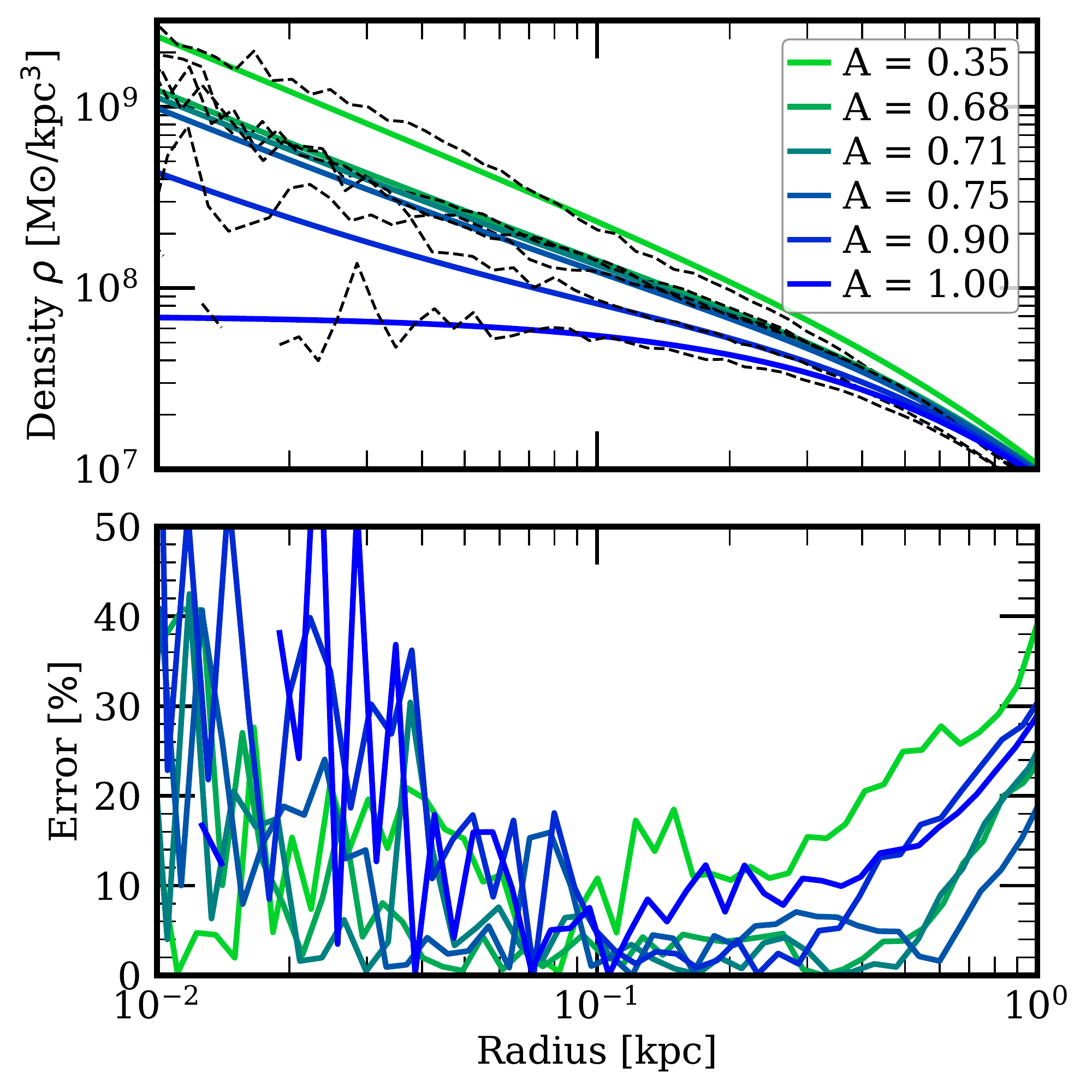}\\
\caption{This plot demonstrates the accuracy of the fitting procedure in \S\ref{interpolation}. The upper panel shows Dehnen potential density profiles with different values of $A$ as fit to the simulation data in R16. The source data is also shown as dashed black lines. The lower panel shows the relative error between the fits and source data, which remain within around 50\% over the relevant range of radii.}
\label{fig:fitting}
\end{figure}

The implementation of dynamical friction in {\sc nbody6df} requires knowledge of the distribution function $f(r,v)$ of the background stars (see \S\ref{df} and \citealt{2015MNRAS.454.3778P}). The distribution function for a Dehnen density profile is fully analytic for $\gamma=0$, $\gamma=1$, and some other key values, but not for arbitrary values of $\gamma$ \citep{1994AJ....107..634T}. For this reason, and in order for our method to be computationally efficient, we implement a time-varying $\gamma(t)$ by performing a linear interpolation between the $\gamma=0$ and $\gamma=1$ scenarios. The background distribution function, which we name $f_m(r,v)$, becomes:
\begin{equation}
f_{m}(r,v) = A \times f_{\rm core} + (1-A)\times f_{\rm cusp},
\label{interpolate.eq}
\end{equation}
where $f_{m}$ is the interpolated distribution function and $A$ is an interpolation variable in the range $A=[0,1]$, corresponding to a cusp for $A=0$ and a core for $A=1$. This simple method is mathematically consistent with a true cusp-core transformation so long as the galactic mass is conserved and the scale radius in equation \ref{dehnen.eq} is suitably interpolated.  
We fixed the scale lengths of the initial and final galactic density profile by fitting the initial and final profile for the $M_{200} = 10^9$\,M$_\odot$ dwarf taken from the simulations in R16. For these fits, the initial $\gamma$ was set to $\gamma=1$, the final $\gamma$ was set to $\gamma=0$, and the mass of the background galaxy was held constant. The interpolation variable $A$ was then fit to the data from R16 such that the resulting interpolation provided a cusp-core transformation matching the time-evolution of the density profile as shown in Fig. \ref{fig:fitting}. The fitting precision was focused around radii within the GC orbit ($r<700$\,pc).

All simulations were run for a Hubble time ($t_{\rm univ} = 14\,$Gyr) or until GC destruction.

\section{Results} \label{sec:results}

\subsection{Visual impression}

Fig. \ref{fig:vis} provides a visual impression of our 128k simulation suite. The panels show mass (top) and luminosity (bottom) weighted maps of the total column density of stars at the end of the simulations, as marked. From left to right, we show results for the cusped ($\gamma=1$), intermediate ($\gamma=0.5$), cored ($\gamma=0$) and time-varying ($\gamma(t)$) simulations. The lower row of each set of panels shows the entire GC orbital plane. The location of the GC density centre is marked by the black spot; the location of the host galaxy centre is marked with a white star. The upper row of each set of panels shows a $\times 10$ zoom closeup of the main GC body, where the dotted white circle marks the initial tidal radius $r_{\rm t}$ (as in equation \ref{tidal.eq}) at $t=0$\,Gyr, and the solid white circle marks the tidal radius at $t=14$\,Gyr.

Firstly, notice that the GC physically grows in size and the tidal debris becomes less prominent as we move from cusped towards cored host potentials. The time-varying potential results in a large GC, but with substantially more tidal debris than in the static cored simulation (compare the right two columns). This occurs because the debris is torn off early in the simulation when the background potential was more cusped, while the GC grows to a larger size as the cusp transforms to a core and the tidal field diminishes. This mis-match between the size of the GC and its tidal debris is, then, a key test for the presence or absence of an historic cusp-core transformation.

We now look more quantitatively at our full simulation suite to study the orbital decay of the GCs orbiting in their host galaxies (\S\ref{sec:orb}), the relaxation times of the GCs (\S\ref{sec:relax}), the tidal tails of the GCs (\S\ref{sec:tails}), the mass-to-light ratio of the GCs within $R_{\rm eff}^\mathcal{M}$ (\S\ref{sec:mlratio}), the sizes of the GCs (\S\ref{sec:hmr}), the velocity dispersion of the GCs within $R_{\rm eff}^\mathcal{M}$ (\S\ref{sec:vdisp}) and the mass segregation of the GCs (\S\ref{sec:segre}). With each of these GC properties, we ask whether we can differentiate between the GC evolving in the $\gamma=0$ galaxy, the $\gamma=1$ galaxy and the $\gamma(t)$ galaxy. This is a {\it minimum} requirement for using GCs to determine their host galaxy potentials. In reality the initial orbit, density and mass of each GC is unknown. These will induce further degeneracies between the different models. We discuss this further in \S\ref{sec:discussion}. \par

\subsection{Orbital radii} \label{sec:orb}
\begin{figure}
\setlength\tabcolsep{2pt}%
\includegraphics[ trim={0 1cm 0 0}, width=\columnwidth, height=\columnwidth, keepaspectratio]{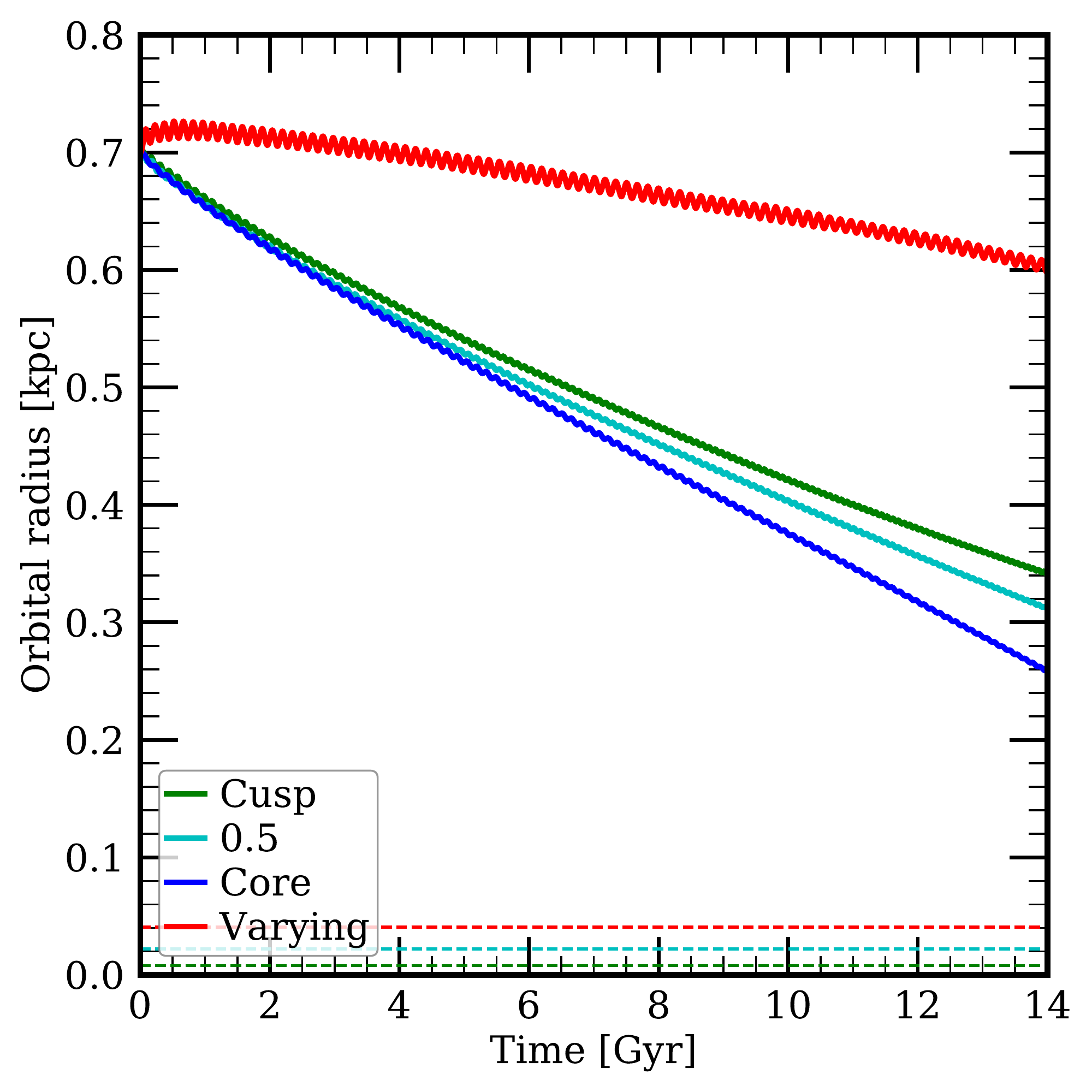}\\
\caption{Orbital radii with time for each of the 128k GC simulations. Small sinusoidal fluctuations are indicative of slightly elliptical orbits despite the care taken to ensure circular orbits. Dotted lines mark the expected stalling-radius, estimated as $r_{\rm stall} = (M_{\rm GC}/M_{\rm g})  ( r_{s}^{2-\gamma} + r_{s}\gamma  )^{1/(3-\gamma)}$.} \label{fig:orbr}
\end{figure}

Plots of the orbital radii for the 128k simulations are shown in Fig. \ref{fig:orbr}. In each simulation, the GC orbit is expected to decay over time due to dynamical friction physics, as described in \S\ref{df}. The 128kcusp, 128k0.5 and 128kcore GCs all infall by a similar amount, with slight differences that can be attributed to their unique mass losses. As will be seen in \S\ref{sec:tails}, galactic potentials with increasing $\gamma$ induce greater tidal stripping, resulting in a less massive GC and less dynamical friction (equation \ref{dynfric.eq}). \par 

Initially, it may seem odd that the 128kcore GC infalls more quickly than the 128kcusp GC prior to mass loss (at $t=0$), given previous work on dynamical friction core-stalling \citep[e.g.][]{2006MNRAS.373.1451R,Inoue09,2018ApJ...868..134K}. However, the result is less surprising after considering Chandrasekhar's formula for dynamical friction in equation~(\ref{dynfric.eq}), which reveals that at $t=0$ we expect a $~10\%$ higher ${{\rm d}v}/{\rm d}t$ for our cored model than for our cusped model. This occurs due to the the satellite velocity, background density and galactic velocity dispersions present. The 128kcore GC begins with a lower circular velocity and therefore a lower orbital energy, and so dynamical friction has a proportionally greater impact on the orbital decay. \par

The 128kvarying GC follows a very different trend, with the orbital radius \textit{growing} during the first Gyr. This is because the GC is pushed out as the galactic potential profile is flattened and the mass internal to the GC orbit drops, acting against dynamical friction. \par

\subsection{Relaxation time} \label{sec:relax}

\begin{table}
\captionof{table}{The number of relaxation times passed within various time intervals for each simulation. Some values are unlisted because the GC has decayed prior to a Hubble time.}
\begin{tabularx}{\linewidth}{lccc} 
 \toprule
 Name & total $t_{\rm relax}$ & total $t_{\rm relax}$ & total $t_{\rm relax}$ \\
  & $t>0$\,Gyr & $0 \leq t \leq 8$\,Gyr & $t>8$\,Gyr \\
 \midrule
 128kcusp     & 9.469 & 6.033 & 3.436 \\ 
 128k0.5      & 7.445 & 5.379 & 2.066 \\
 128kcore     & 6.128 & 4.789 & 1.339 \\ 
 128kvarying  & 8.245 & 6.082 & 2.163 \medskip  \\
 64kcusp      & 15.974 & 7.264 & 8.710 \\
 64k0.5       & 10.397 & 6.786 & 3.611 \\
 64kcore      & 7.520 & 5.754 & 1.766 \\
 64kvarying   & 8.900 & 6.525 & 2.375 \medskip  \\
 32kcusp      & -     & 10.381 & -     \\
 32k0.5       & -     & 6.429 & -     \\
 32kcore      & 9.969 & 6.965 & 3.004 \\
 32kvarying   & 12.469 & 8.241 & 4.228 \\
 \midrule
 64kcusplow   & -     & 4.353 & -     \\
 64k0.5low    & 6.077 & 3.667 & 2.410 \\
 64kcorelow   & 5.019 & 3.688 & 1.331 \\
 64kvaryinglow& 5.320 & 3.826 & 1.494 \medskip  \\
 64kcusphigh    & 21.859 & 13.525 & 8.334 \\ 
 64k0.5high     & 16.189 & 10.891 & 5.298 \\
 64kcorehigh    & 11.394 & 8.966 & 2.428 \\
 64kvaryinghigh & 14.481 & 11.210 & 3.271 \\
 
\end{tabularx}
\label{tab:trelax}
\end{table}

The half-mass relaxation time can be approximated with \citep{1971ApJ...164..399S}:
% http://adsabs.harvard.edu/abs/1969ApJ...158L.139S
\noindent\begin{equation}
	t_{\rm relax} = \frac{0.138 \left ( N R_{1/2}^{3} / Gm \right )}{\log (\gamma N)},
	\label{relax.eq}
\end{equation}
where $N$ is the number of particles within $R_{1/2}$, $m$ is the average mass of these particles and $\gamma=0.02$ for multimass systems \citep{1996MNRAS.279.1037G}. \par

Table \ref{tab:trelax} lists the number of relaxation times (with the relaxation time recalculated for each time step) passed for each simulation within time intervals relevant to the cusp-core transformation time $t_{\rm tr}$ as in \S\ref{sec:intro}. These results are important to consider in addition to the initial $t_{\rm relax}$ from Table \ref{tab:sim1} because $t_{\rm relax}$ varies significantly with the GC evolution. \par

The results here are consistent with the observations we will see later in \S\ref{sec:hmr} in that the sets of simulations at lower particle numbers elapse more relaxation times despite their lower stellar density. In addition, there is a common pattern whereby GCs in $\gamma=1$ undergo more relaxation times, although the majority of this effect occurs after $t_{\rm tr}$ when the GC is exposed to stronger tides due to its orbital decay and, as a result, has begun to dissolve. \par

\subsection{Tidal tails} \label{sec:tails}

\begin{figure*}
  \centering
  \setlength\tabcolsep{2pt}%
  \begin{tabularx}{\textwidth}{c}
    {\Large Mass-weighted} \\
    \includegraphics[trim={1cm 0cm 0cm 0cm}, width=\linewidth]{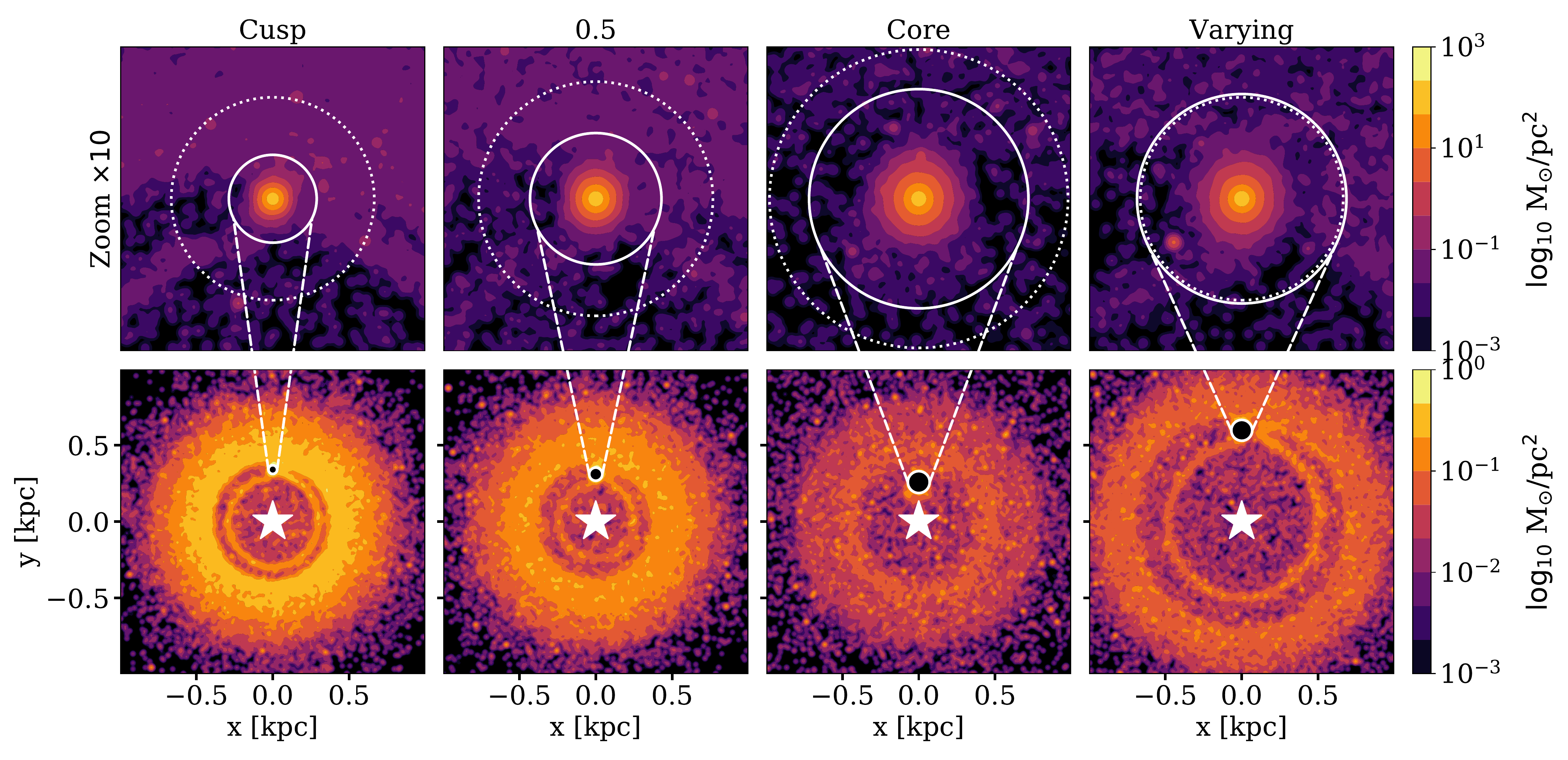} \\
    {\Large Luminosity-weighted} \\
    \includegraphics[trim={1cm 0cm 0cm 0cm}, width=\linewidth]{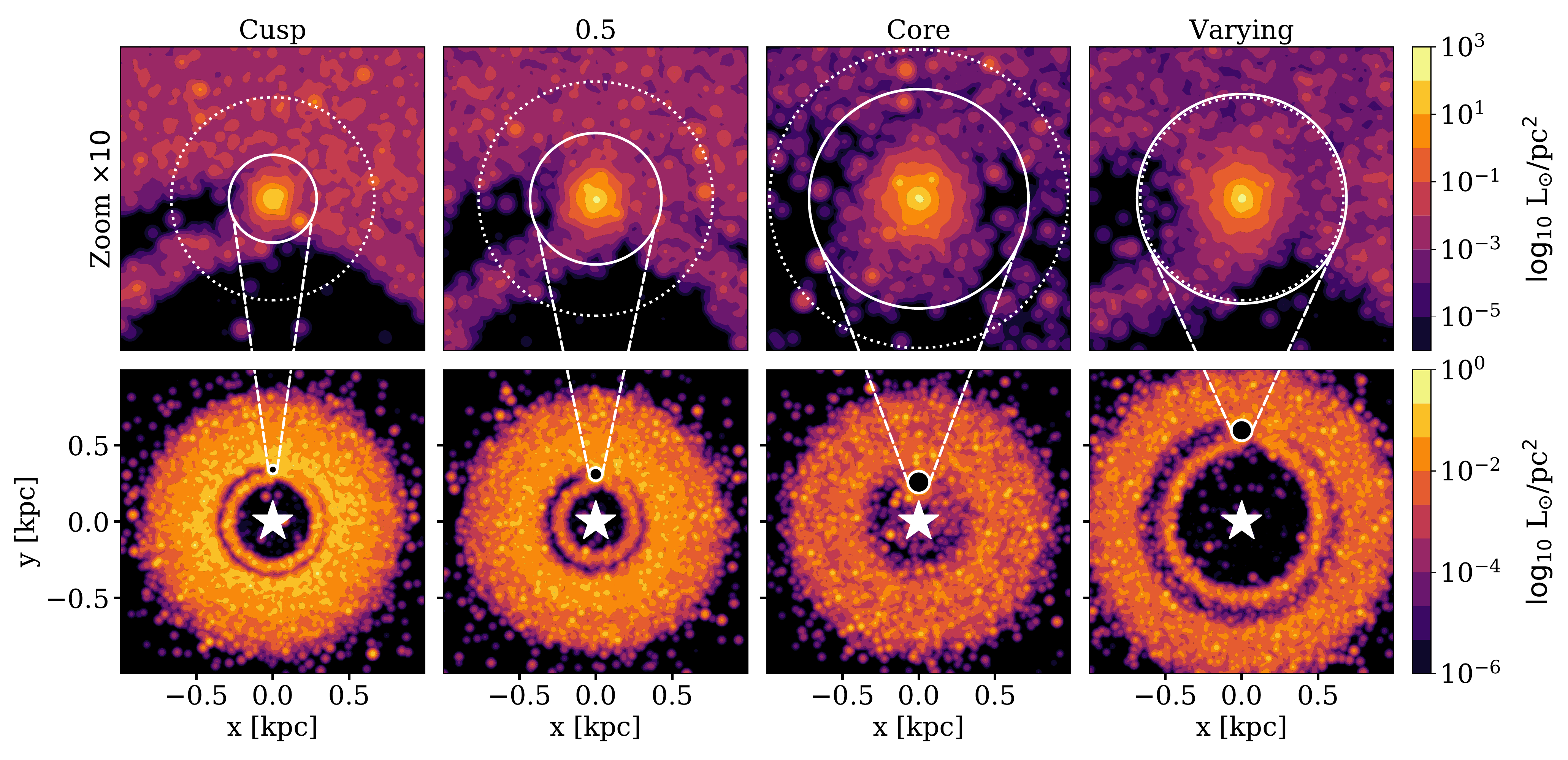} \\
  \end{tabularx}
\caption{Mass- (top) and luminosity- (bottom) weighted maps of the column density of each 128k simulated GC after a Hubble time, as marked. From left to right, the panels show results for the cusped ($\gamma=1$), intermediate  ($\gamma=0.5$), cored ($\gamma=0$) and time-varying ($\gamma(t)$) simulations. The lower row of each set of panels shows the entire GC orbital plane. The location of the GC density centre is marked by a black spot; the location of the host galaxy centre is marked with a white star. The upper row of each set of panels shows a $\times 10$ zoom closeup of the main GC body, where the dotted white circle marks the initial tidal radius $r_{\rm t}$ as in equation \ref{tidal.eq} at $t=0$\,Gyr, and the solid white circle marks the tidal radius at $t=14$\,Gyr. The tidal radii at $t=14$\,Gyr are equal to 28.76, 43.02, 71.79 and 68.56\,pc for the cusp, 0.5, core and varying panels respectively.} \label{fig:vis}
\end{figure*}

In Fig. \ref{fig:vis}, we give a visual representation of each of our 128k GC simulations. The panels show mass and luminosity-weighted column density maps, as marked. The lower row of each set of panels shows the entire GC orbital plane. The location of the GC density centre is marked by the black spot; the location of the host galaxy centre is marked with a white star. The upper row of each set of panels shows a $\times 10$ zoom closeup of the main GC body, where the white circle marks the tidal radius $r_{\rm t}$. The tidal radius \citep[equation 7 in][]{1962AJ.....67..471K} for a circular orbit in a Dehnen profile can be calculated with \citep*[equation 10 in][]{2011MNRAS.418..759R} as:
\noindent\begin{equation} \label{tidal.eq}
  r_{\rm t} = \left [ \frac{M_{\rm GC}}{M_{\rm g}}
  \frac{r^\gamma\left(r_{\rm s}+r \right )^{4-\gamma} }{3r+r_{\rm s}\gamma} \right ] ^{1/3}.
\end{equation}
\par

We can then approximate the `Roche volume' that encompasses stars that remain bound to the GC, as $V_{\rm R} = (4/3) \pi r_{\rm t}^3$.
Notice that the tidal tails formed from tidally stripped stars are clearly visible in the lower panels, constructing a connected ring around the host galaxy. The tidal tails are denser with increasing host galaxy DM cusp slope, $\gamma$, due to the stronger gravitational tides of galactic potential with increasing $\gamma$. The 128kvarying simulation produces tidal tails that are slightly denser than that of 128kcore. This owes in part to the more extended orbital radius of the 128kvarying GC and the associated weaker tides (see \S\ref{sec:orb}). Additionally, the orbital radius of 128kvarying has occupied a narrower region of space throughout simulation time (Fig. \ref{fig:orbr}). As a consequence, the tidal debris has been shed within a narrower annulus which increases the tidal tail density. \par 

The total mass of stars in the tidal tails of each simulation, estimated as the total mass excluding the mass within $V_{\rm R}$ and particles exterior to twice the initial GC orbital radius, is summarised in Table \ref{tab:tailmass}. This estimate would suggest it is possible to differentiate between the simulations based on tidal tail mass. It should be noted that elliptical orbits would complicate our ability to differentiate between models; we will consider this in future work.
\par 

\begin{table}
\centering
\captionof{table}{Total tidal tail mass for each 128k simulation, where $M_{\rm tails}$ is the total mass of particles excluding those within $V_{\rm R}$ and exterior to twice the initial GC orbital radius.}
\begin{tabular}{lc} 
 \toprule
 Name & $M_{\rm tails}$ (M$_{\odot}$) \\
 \midrule
 128kcusp     & $3.570 \times 10^4$ \\ 
 128k0.5      & $2.569 \times 10^4$ \\
 128kcore     & $1.254 \times 10^4$ \\ 
 128kvarying  & $2.059 \times 10^4$ \\
 \bottomrule
\end{tabular}
\label{tab:tailmass}
\end{table}

\subsection{Changing the GC mass and density: effect on the evolution of the half-mass and half -light radii} \label{sec:hmr}

% R1/2 and L1/2 plots for 128k, 64k and 32k GCs
\begin{figure}
\setlength\tabcolsep{2pt}%
\includegraphics[ trim={0.9cm 0.3cm 0.3cm 0.3cm}, width=\columnwidth]{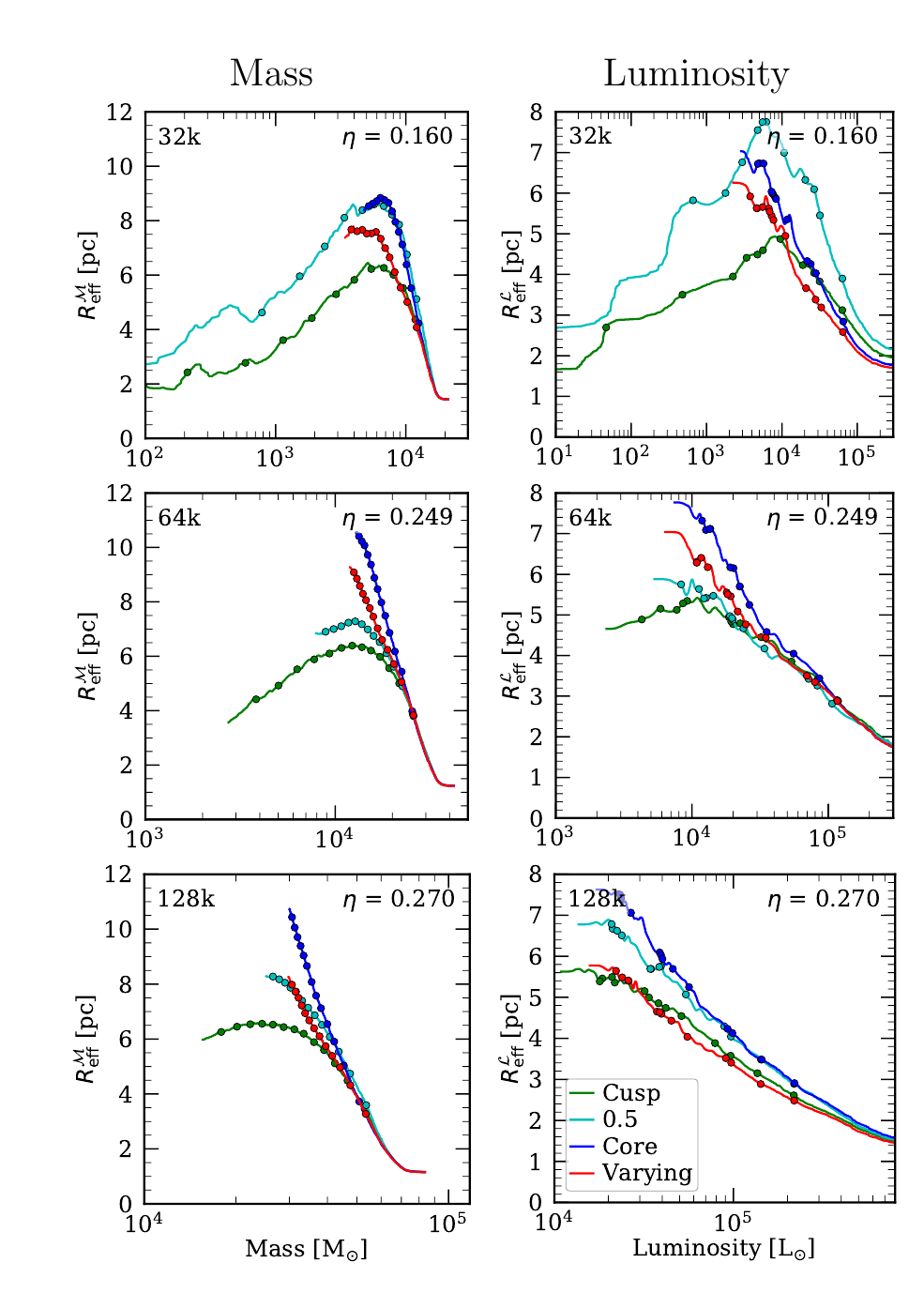}
\caption{Plots of the  GCs' $R_{\rm eff}^\mathcal{M}$ (left panels) and $R_{\rm eff}^\mathcal{L}$ (right panels) as a function of the total mass and luminosity of stars within the tidal radius $r_{\rm t}$ (Equation \ref{tidal.eq}). for the 32k, 64k and 128k simulations. The circles indicate 1\,Gyr time intervals, with the arrow of time pointing from right to left. The number in the top left corner of each panel represents the GC size, whereas the number in the top right corner represents the ratio of the average relaxation time to the cusp-core transformation time: $\eta = \left \langle t_{\rm relax} \right \rangle /t_{\rm tr}$ where $t_{\rm tr}$ is given in \S\ref{sec:intro}. The results have been smoothed with a Gaussian filter.} \label{fig:128krh}
\end{figure}

% R/12 and L1/2 plots for 64k GCs at different densities
\begin{figure}
\setlength\tabcolsep{2pt}%
\includegraphics[ trim={0.9cm 0.3cm 0.3cm 0.3cm}, width=\columnwidth]{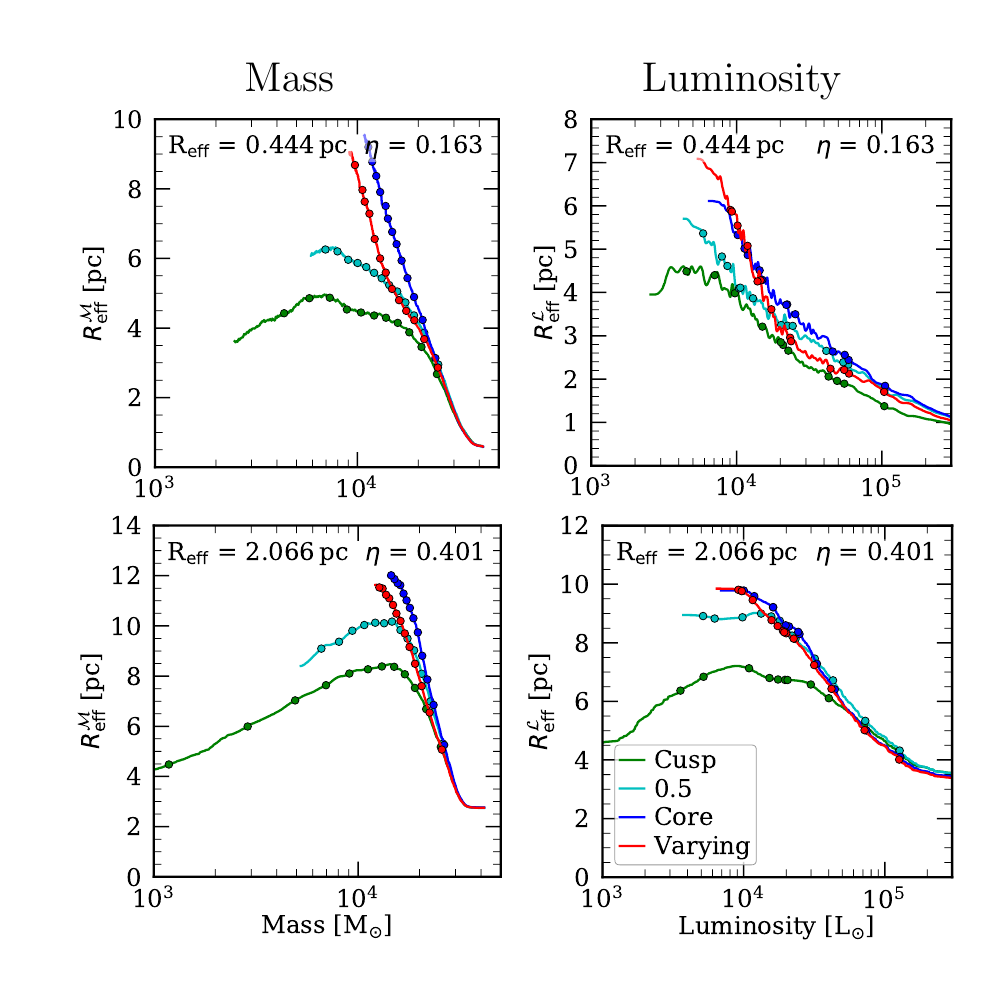}
\caption{As Fig. \ref{fig:128krh} but for the 64k simulations with different initial GC densities, as specified in Table \ref{tab:sim1}. The top panels correspond to the 64khigh simulations, while the bottom panels correspond to the 64klow simulations.} \label{fig:64krh}
\end{figure}

The size evolution of the GCs can be summarised by plotting the projected half-mass radius $R_{\rm eff}^\mathcal{M}$ and projected half-light radius $R_{\rm eff}^\mathcal{L}$ against the mass within $V_{\rm R}$. GC mass provides a good analogue for evolution time because GCs decay continuously due to stellar evaporation and tidal stripping \citep{bandt}. 
Although there are no velocity kicks for BHs, only a fraction are retained within $2 \times R_{\rm eff}^\mathcal{M}$ by the end of the simulation. Of an initial 255 BHs for the 128k simulations, 65, 78, 78 and 46 are retained for the GCs in a $\gamma = 1, 0.5, 0$ and $\gamma(t)$ potential respectively.
Fig. \ref{fig:128krh} shows these plots for the four main simulation suites introduced in Table \ref{tab:sim1}. \par

The trends for both 128k and 64k simulations demonstrate a clear difference in the evolution of $R_{\rm eff}^\mathcal{M}$ and $R_{\rm eff}^\mathcal{L}$ depending on $\gamma$. The $R_{\rm eff}^\mathcal{M}$ growth of the 128kcusp and 64kcusp simulations begins to flatten off within a few Gyr as a result of the clusters approaching their minimum densities set by the tides, whereas those of 128kcore and 64kcore grow almost linearly, far surpassing their counterparts evolving in a $\gamma=1$ galaxy by a Hubble time. The behaviors of 128k0.5 and 64k0.5 are an intermediate between that of the $\gamma=0$ and $\gamma=1$ simulations. Notice that the evolution of $R_{\rm eff}^\mathcal{M}$ in the $\gamma(t)$ simulations are distinct from that of the $\gamma=0$ simulations after a few Gyr. We will return to this shortly.\par

The results for the 32k simulations in Fig. \ref{fig:128krh} are a slight oddity, whereby the trend for 32k0.5 is difficult to describe as intermediate between the 32kcusp and 32kcore simulations. This is likely due to numerical noise. Regardless, the trend for 32kvarying is such that a smaller $R_{\rm eff}^\mathcal{M}$ is yielded after a Hubble time when compared to 32kcore, in accordance with the results from the 128k and 64k simulations. \par

The mass versus $R_{\rm eff}^\mathcal{M}$ trend for simulations with different $N$ mature at different rates, as expected given the different initial relaxation times in Table \ref{tab:sim1} (and see also the total elapsed relaxation times in \S\ref{sec:relax}). Assuming the GCs were allowed unlimited evolution time, all simulations should reach a maximum $R_{\rm eff}^\mathcal{M}$ and then decay. In the case of 128kcore there is not yet any indication of a plateau, which is due to its slower dynamical evolution. \par

The principal behaviors observed in Fig. \ref{fig:128krh} are reproduced by the simulations of GCs with varying initial densities, shown in Fig. \ref{fig:64krh}. In these simulations, the difference between the $\gamma=0$ and $\gamma(t)$ simulations are less distinct, with the differences in $R_{\rm eff}^\mathcal{L}$ being difficult to distinguish after a Hubble time for the 64k0.1 simulations. It should also be noted that the final $R_{\rm eff}^\mathcal{L}$ has some dependence on the GC density, and this would have to be taken into account when making comparisons with observations. \par

The results for the 64kvarying0.1 simulation can be explained by the higher $t_{\rm relax}$ at the low initial GC density (Table \ref{tab:sim1}). With a higher $t_{\rm relax}$ to $t_{\rm tr}$ (\S\ref{sec:intro}) ratio, the GCs have not had sufficient time to adapt to their respective potentials before the cusp-core transformation is well underway. This means the GC in simulation 64kvarying0.1 has had little opportunity to adopt the $R_{\rm eff}^\mathcal{M}$ trend associated with $\gamma=1$. The opposite is the case for the 64kvarying10 simulation, where the high initial GC density leads to a lower $t_{\rm relax}$. Here the GC adapts to its host potential so rapidly that the cusp-core transformation has created an `elbow' which is visually apparent in the $R_{\rm eff}^\mathcal{M}$ trend of 64kvarying10. \par

The $R_{\rm eff}^\mathcal{L}$ trends exhibit much the same results, with a few oddities: In the 128k simulations the order of the $\gamma=1$ and $\gamma(t)$ trend has reversed. Fig. \ref{fig:mlratio} shows how this is a consequence of the lower mass-to-light ratio of the GC in 128kcusp. \par

\subsection{Mass-to-light ratio} \label{sec:mlratio}

% M/L ratios
\begin{figure}
\setlength\tabcolsep{2pt}%
   \includegraphics[ width=\columnwidth, keepaspectratio, trim={0cm 1.5cm 0cm 0cm}]{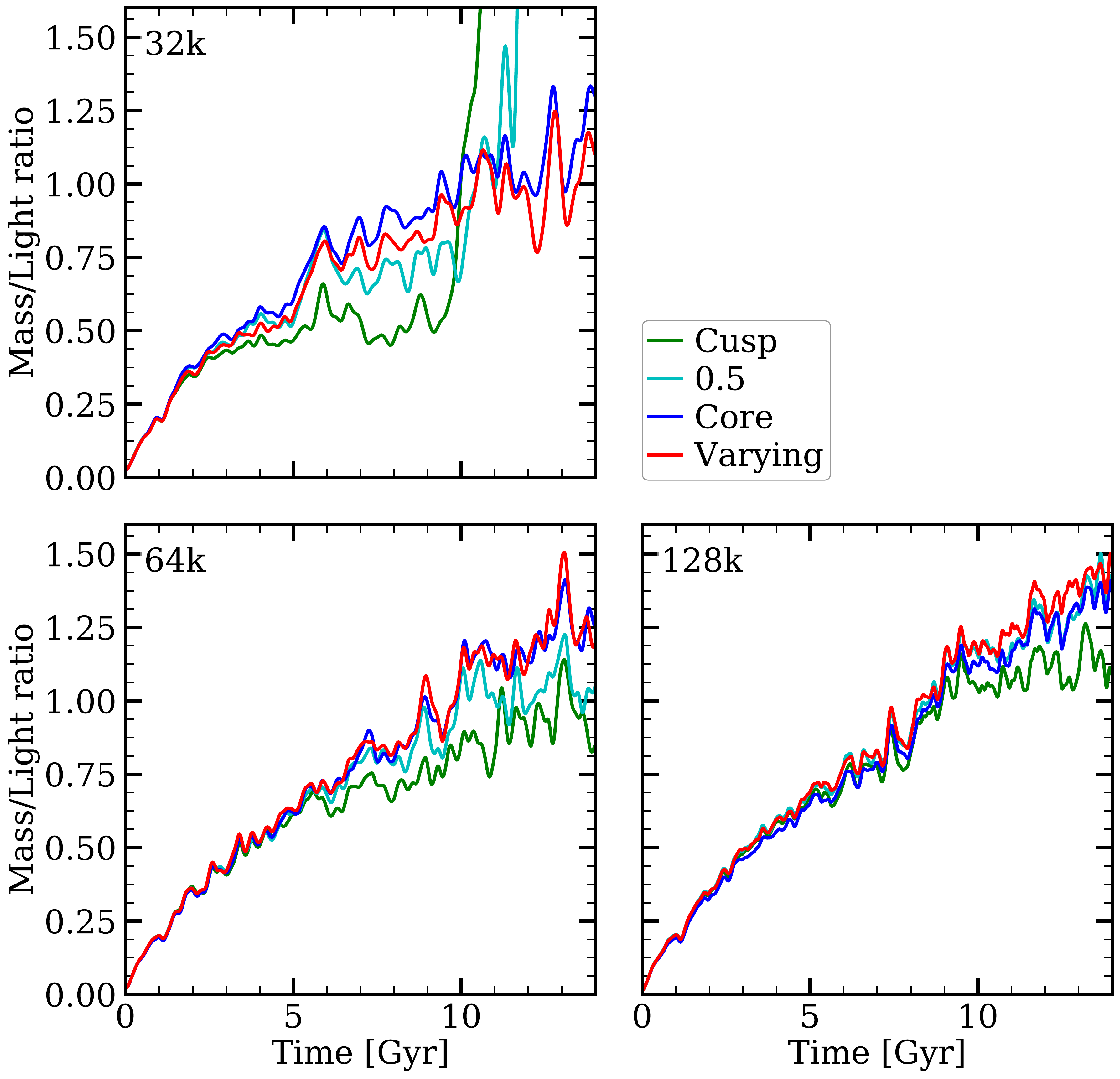} \\
\caption{Plots of the stellar mass-to-light ratio within $r_{\rm t}$ (Equation \ref{tidal.eq}) with time. for the 32k, 64k and 128k simulations. Results have been smoothed with a Gaussian filter.}
\label{fig:mlratio}
\end{figure}
The stellar mass-to-light ratios for each GC as a function of time are shown in Fig. \ref{fig:mlratio}. These show an increase in mass-to-light ratios for all GCs, which is the expected consequence of stellar evolution which results in stellar luminosity reducing more so than stellar mass. \par 

The relationships in each panel of Fig. \ref{fig:mlratio} diverge after a few Gyr such that there is a slightly lower mass-to-light ratio with increasing host galaxy DM cusp slope. The onset of this divergence is linked to the number of particles $N$, with the larger simulations taking significantly longer to diverge. This is due to the more massive GCs having a correspondingly higher $t_{\rm relax}$, and we explore this in \S\ref{sec:relax}. The slower rise of the mass-to-light ratio in GCs with increasing host galaxy DM cusp slope is a result of the preferential ejection of low-mass stars with high mass-to-light ratios, which is due to mass segregation which we discuss in \S\ref{sec:segre}. Whilst this effect occurs in all of the GCs, it is proportionally more significant in the GCs with higher mass loss rates. \par

The GCs in $\gamma(t)$ maintain a mass-to-light ratio almost identical to that of the GCs in $\gamma=0$ at all times. This is because the host galaxy potential is mostly cored ($\gamma < 0.5$) by 8\,Gyr, before the mass-to-light ratios for the GCs in static background potentials diverge. \par

The results for 32kcusp and 32k0.5 are skewed after $\sim 10$\,Gyr because the GC body evaporates. When a GC evaporates it goes through a brief phase where it is dominated by dark stellar remnants, resulting in a sharp spike in the mass-to-light ratio \citep*{2009A&A...502..817A}. With the exception of these two simulations, there is a consistent pattern across each panel. \par

\subsection{Velocity dispersion} \label{sec:vdisp}

% Velocity dispersion of 128k GCs after a Hubble time
\begin{figure}
\setlength\tabcolsep{2pt}%
\includegraphics[ trim={0 1cm 0 0}, width=\columnwidth, height=\columnwidth, keepaspectratio]{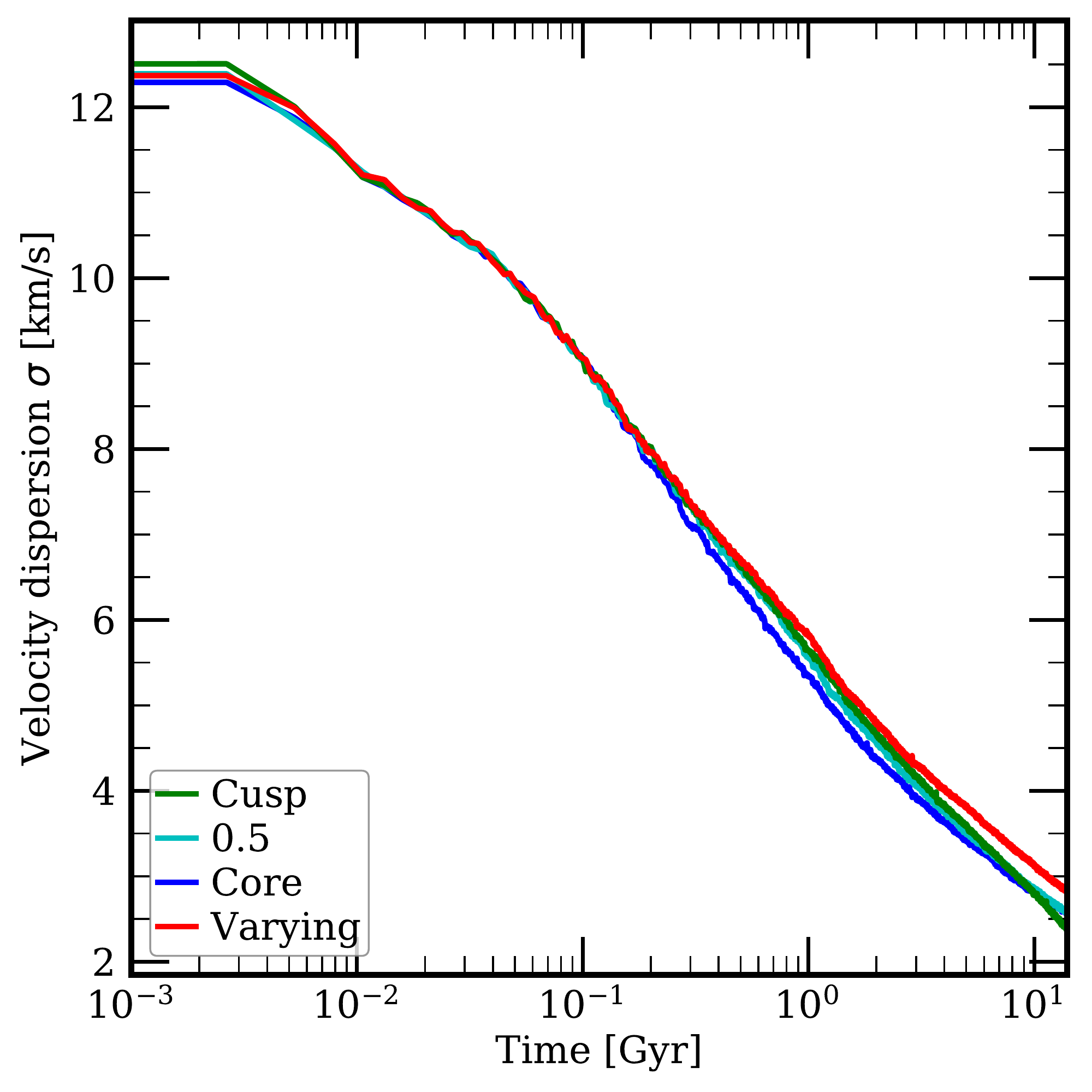}\\
\caption{Plot showing the velocity dispersion within $R_{\rm eff}^\mathcal{M}$ for GCs in 128k simulations as a function of time. Trends have been normalised to the 128kcore simulation to expose the small differences between each simulation.} \label{fig:vdisp}
\end{figure}

Assuming a virialised cluster, the GC velocity dispersion can be shown to vary with the mass and radius as summarised in equation \ref{sigma.eq}: 
\noindent\begin{equation}
\label{sigma.eq}
\sigma \sim \sqrt{\frac{GM}{R_{1/2}}}
\end{equation}
From Fig. \ref{fig:128krh} we see that over time the GC mass tends to decrease whilst $R_{\rm eff}^\mathcal{M}$ tends to increase. Therefore it is expected that the GC velocity dispersion should decrease with time. \par

We calculate the velocity dispersion $\sigma$ with:
\noindent\begin{equation}
\sigma = \sqrt{\frac{\sum_{i=0}^{N}(\upsilon_{i}-\bar{\upsilon} )^{2}}{N}},
\label{vdisp.eq}
\end{equation} 
where $\upsilon_i$ is a particle's velocity and $N$ is the number of particles within $R_{\rm eff}^\mathcal{M}$. \par

In Fig. \ref{fig:vdisp}, we plot the total velocity dispersion of particles within $R_{\rm eff}^\mathcal{M}$ for GCs in 128k simulations as a function of time. For all times, there is hardly any difference between GCs orbiting different host galactic mass profiles. It is therefore unlikely that the observed velocity dispersions of GCs could be used to distinguish between different host galaxy galactic potentials. \par

\subsection{Mass segregation} \label{sec:segre}

% Mass segregations of 128k GCs at significant times
\begin{figure}
\setlength\tabcolsep{2pt}%
\includegraphics[ width=\columnwidth, keepaspectratio]{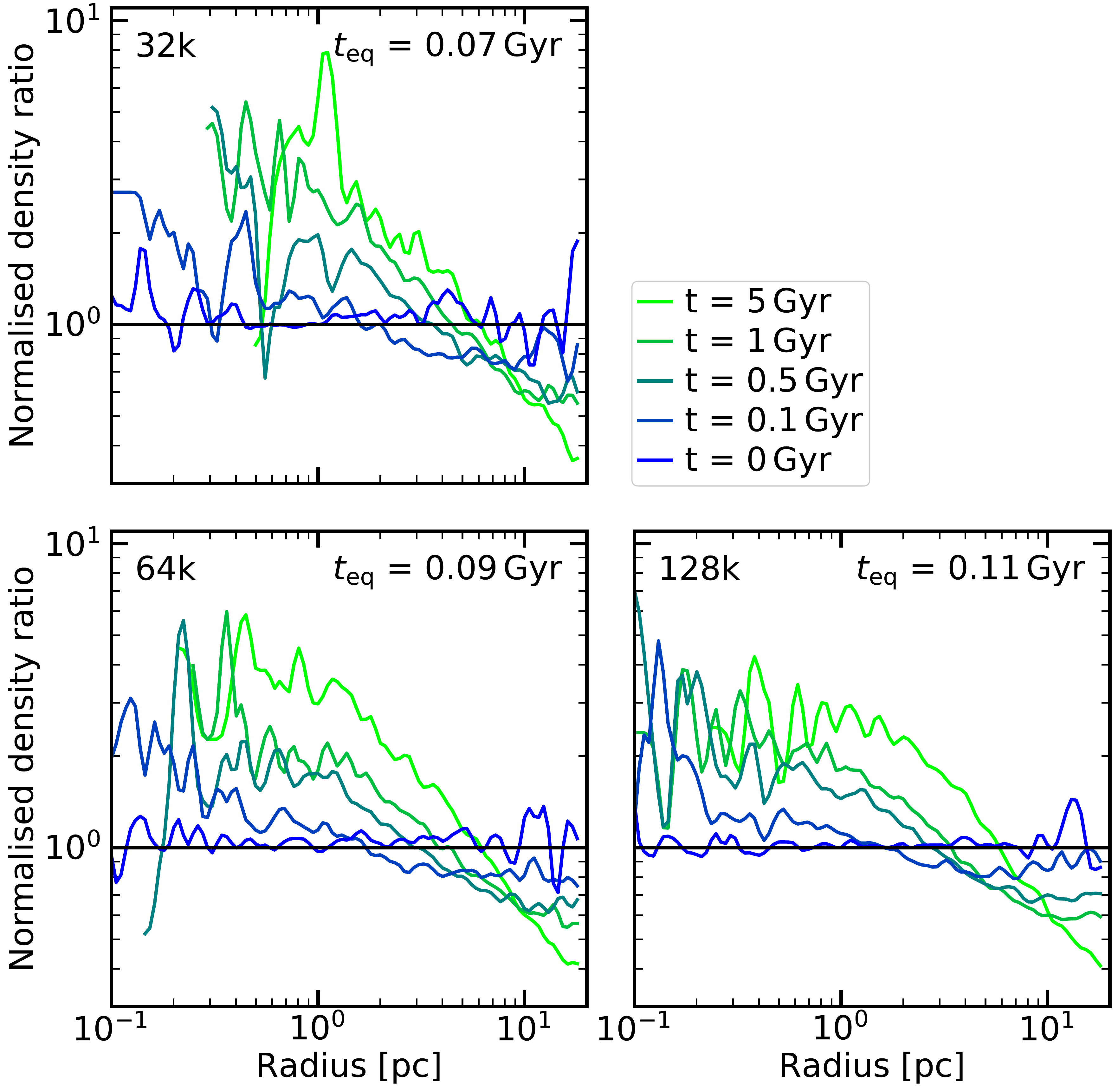}
\caption{Plots showing mass segregation for GCs in the 32k, 64k and 128k simulations. The mass segregation is calculated over five different time intervals, with blue lines representing earlier times and green lines representing later times. The mass segregation was found by grouping stars within $V_{\rm R}$ into two mass bins: $M_{\rm low} = 0.5-1.0$\,M$_{\odot}$ and $M_{\rm high} = 1.0-200.0$\,M$_{\odot}$, and plotting the normalised density ratio of the two, $\zeta(r,t)$ (Equation \ref{zeta.eq}) as a function of radius. Mass segregation causes $\zeta$ to rise at small radii, indicating an inward movement of massive stars, and $\zeta$ to fall at large radii, indicating an outward movement of low-mass stars. An approximation of the equipartition timescale (Equation \ref{teq.eq}) is included in the top right corner of each panel. A black horizontal line marks the $\zeta=1$ point. The radial ranges are as large as the data reasonably allowed. The results have been smoothed with a Gaussian filter.} \label{fig:segre}
\end{figure}

GC structure such as the mass segregation of the stellar profile has been well observed in nature and is anticipated as a result of equipartition of energy \citep{1996ASPC...92..257A}. A possibility considered here is that the development of the GC mass segregation is influenced by the $\gamma$ of the host galaxy, which may lead to unique GC properties. \par

To measure the amount of mass segregation, we grouped all stars within the $V_{\rm R}$ into two mass bins: $M_{\rm low} = 0.1-0.5$\,M$_{\odot}$ and $M_{\rm high} = 0.5-2.0$\,M$_{\odot}$. We then defined a `normalised density ratio':
\begin{equation}
\zeta(r,t) = \frac{N_{\rm high}(r,t) / N_{\rm low}(r,t)}{N_{\rm high}(R_{\rm eff}^\mathcal{M},t) / N_{\rm low}(R_{\rm eff}^\mathcal{M},t)}
\label{zeta.eq}
\end{equation}
where $N_{\rm low}(r,t)$ is the number of stars in the mass bin $M_{\rm low}$ within radius $r$ and at time $t$, and similarly for $N_{\rm high}(r,t)$. The ratio is normalised such that $\zeta=1$ at $R_{\rm eff}^\mathcal{M}$.
In Fig. \ref{fig:segre}, we plot $\zeta(r,t)$ for the GCs in galactic profiles with $\gamma=1$ in the main simulation suite. For each simulation we plot $\zeta$ at five different times, with $t=0, 0.1, 0.5, 1, 5$\,Gyr. Included in the upper right corner is the equipartition timescale $t_{\rm eq}$ as in \citet{1940MNRAS.100..396S, 1962pfig.book.....S}:
\begin{equation}
t_{\rm eq} = \frac{\left ( \left \langle \upsilon_{1}^{2} \right \rangle + \left \langle \upsilon_{2}^{2} \right \rangle \right )^{3/2}}{8(6\pi)^{1/2} \rho_{01} G^{2} m_{2} \ln{N_{1}}}, 
\label{teq.eq}
\end{equation}
where $\upsilon_{1,2}$ refer to the velocities of the particles in the low-mass bin and the high-mass bin respectively, $\rho_{01}$ is the central density of the particles in the low-mass bin, $m_{2}$ is the average particle mass in the high-mass bin and $N_{1}$ is the number of particles in the low-mass bin. This equation was designed for use with bins of two discrete particle masses, whereas here we use bins of varied particle masses. This timescale is independent of any external potential, and is roughly the same between GCs of different initial $N$. The $t_{\rm eq}$ timescales shown in Fig. \ref{fig:segre} are extremely short, so any secondary influences such as mass loss through the GC tidal boundary will be negligible. They are also relatively similar, so we do not expect any major differences in the time taken to become mass segregated.

This may seem surprising because $t_{\rm eq}$ varies with $N$, but consider that the initial radii of our GCs are independent of $N$ (as can be inferred from Table \ref{tab:sim1}) such that $\rho_0$ is proportional to $N$, and it can be seen from equation \ref{teq.eq} that the two parameters act to cancel each other out. \par

All simulations start unsegregated at $t=0$\,Gyr with $\zeta \simeq 1$ over all radii, with deviations at ranges where there is low sampling density. The GCs become increasingly mass segregated as they age dynamically, starting from within $R_{\rm eff}^\mathcal{M}$ and working outwards. This can be seen in the slope of $\zeta$ becoming steeper and steeper with each time step. This is an expected result, as stellar interactions that drive mass segregation occur at higher probabilities in the denser central regions of the GC. By approximately 1\,Gyr all GCs are maximally mass segregated, with little change occurring after 1\,Gyr. The mass segregation in all simulations occurs on approximately the same timescale. \par

As with the velocity dispersion, there is no discernible difference between the simulations with different $\gamma$ and so we do not show these in Fig. \ref{fig:segre}. This is because the majority of mass segregation has already occurred by 1\,Gyr, on a timescale much smaller than $t_{\rm tr}$ (\S\ref{sec:intro}) and before the GC sizes diverge (see Fig. \ref{fig:128krh}). In effect, maximal mass segregation is reached before the GC properties have responded to the galactic potential in which they are orbiting. \par

\section{Discussion}\label{sec:discussion}

\subsection{Comparison with previous work}

\citet{2018MNRAS.479.3708W} have recently conducted a similar numerical study to ours, simulating the properties of star clusters evolving in static cored and static cusped background potentials. The key differences between their simulations and ours are that our 128k GCs are $\sim 4 \times$ more massive and have a $\sim 10 \times$ smaller half-mass radius. They also simulate a host galaxy of roughly half the mass of ours. In good agreement with their findings, we find that GC size evolution is highly dependent on the central slope of the host galaxy. However, our results in Fig. \ref{fig:segre} appear to be in contradiction with the authors' findings. The authors observe unique mass segregation timescales depending on the shape of the host galaxy central slope. However, their lower GC mass and density yield a much higher $t_{\rm eq}$ (Equation \ref{teq.eq}). This means the GCs in \citet{2018MNRAS.479.3708W} will have had much more opportunity to adapt to their host potential prior to becoming maximally mass segregated. \par

Our results are in good agreement with earlier studies that find that GCs in dwarf galaxies are efficiently destroyed by steep dark matter cusps \citep[e.g.][]{2006MNRAS.368.1073G,2012MNRAS.426..601C,2017ApJ...844...64A}. Furthermore, our result linking large GC sizes to the presence of a central dark matter core is also seen in \citet{2018MNRAS.476.3124C} and \citet{2018MNRAS.479.3708W}. The key new result in our paper, however, is that GCs can survive and grow to large sizes also in a dwarf galaxy that has undergone a gradual cusp-core transformation. \par

\subsection{GCs evolve to larger sizes in a cored background potential}

In this section, we compare our results to a selection of observed GCs in the Fornax \citep[Table 11]{2005ApJS..161..304M}, NGC 6822 \citep{2011ApJ...738...58H, 2015MNRAS.452..320V}, IKN (\citealp{2009MNRAS.392..879G}; \citealp*{2015A&A...581A..84T}), SMC \citep{2009AJ....138.1403G} and Sagittarius \citep{1996AJ....112.1487H, 2010arXiv1012.3224H} dwarf galaxies. This selection was chosen due to the high quality observations available for these galaxies. We contrast these with metal-rich GCs in the Milky Way that are likely to have formed in-situ (\citealp{2010MNRAS.404.1203F, 1996AJ....112.1487H, 2010arXiv1012.3224H}; \citealp*{2013MNRAS.436..122L}) and young star clusters currently forming in M83 \citep{2015MNRAS.452..525R}. This latter gives us some handle on the likely distribution of birth-sizes for globular clusters. \par

\begin{figure}
  \setlength\tabcolsep{2pt}%
  \begin{tabularx}{\columnwidth}{c}
    \includegraphics[trim={0cm 0cm 0cm 0cm}, width=\columnwidth, height=\columnwidth, keepaspectratio]{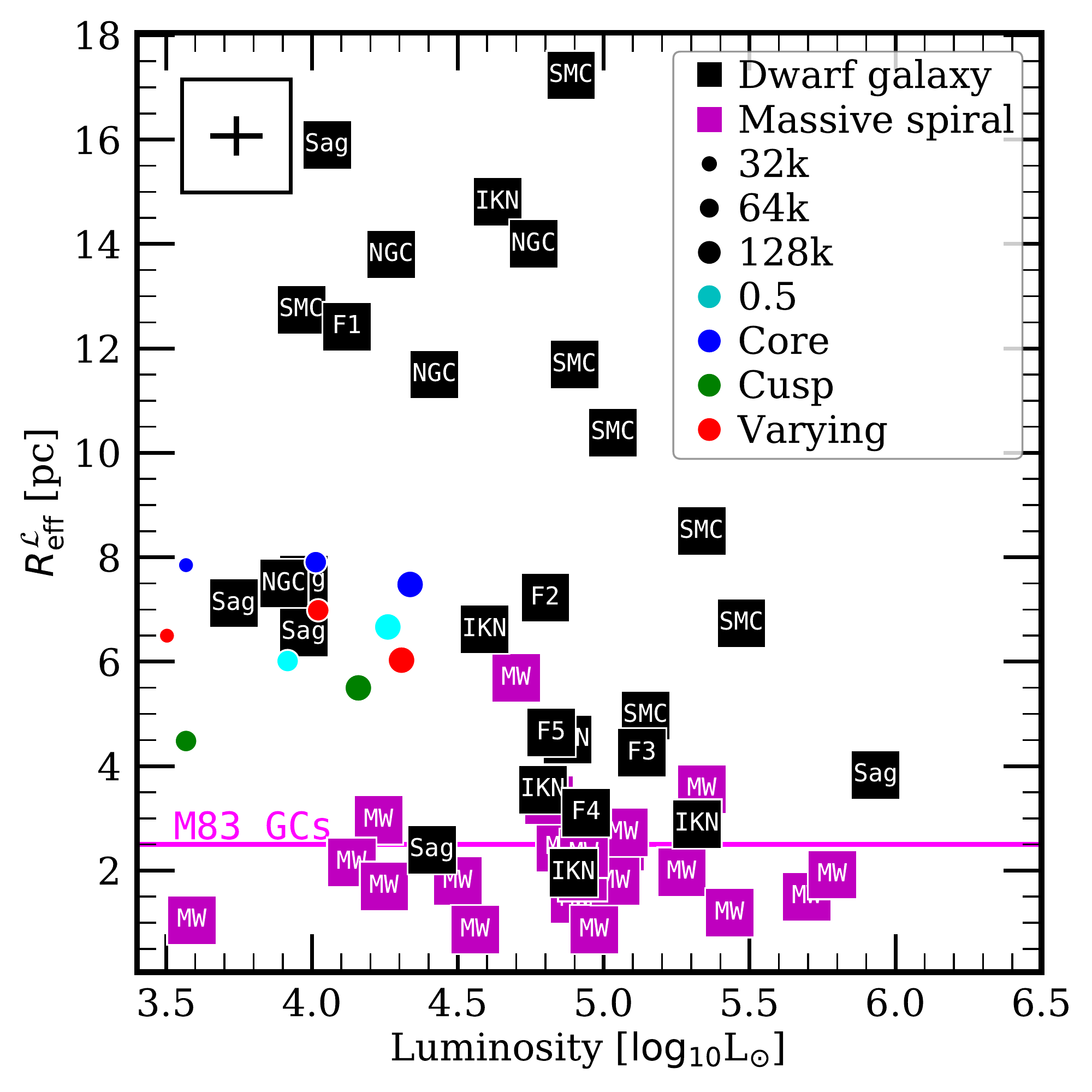} \\
    \includegraphics[trim={0cm 0cm 0cm 0cm}, width=\columnwidth, height=\columnwidth, keepaspectratio]{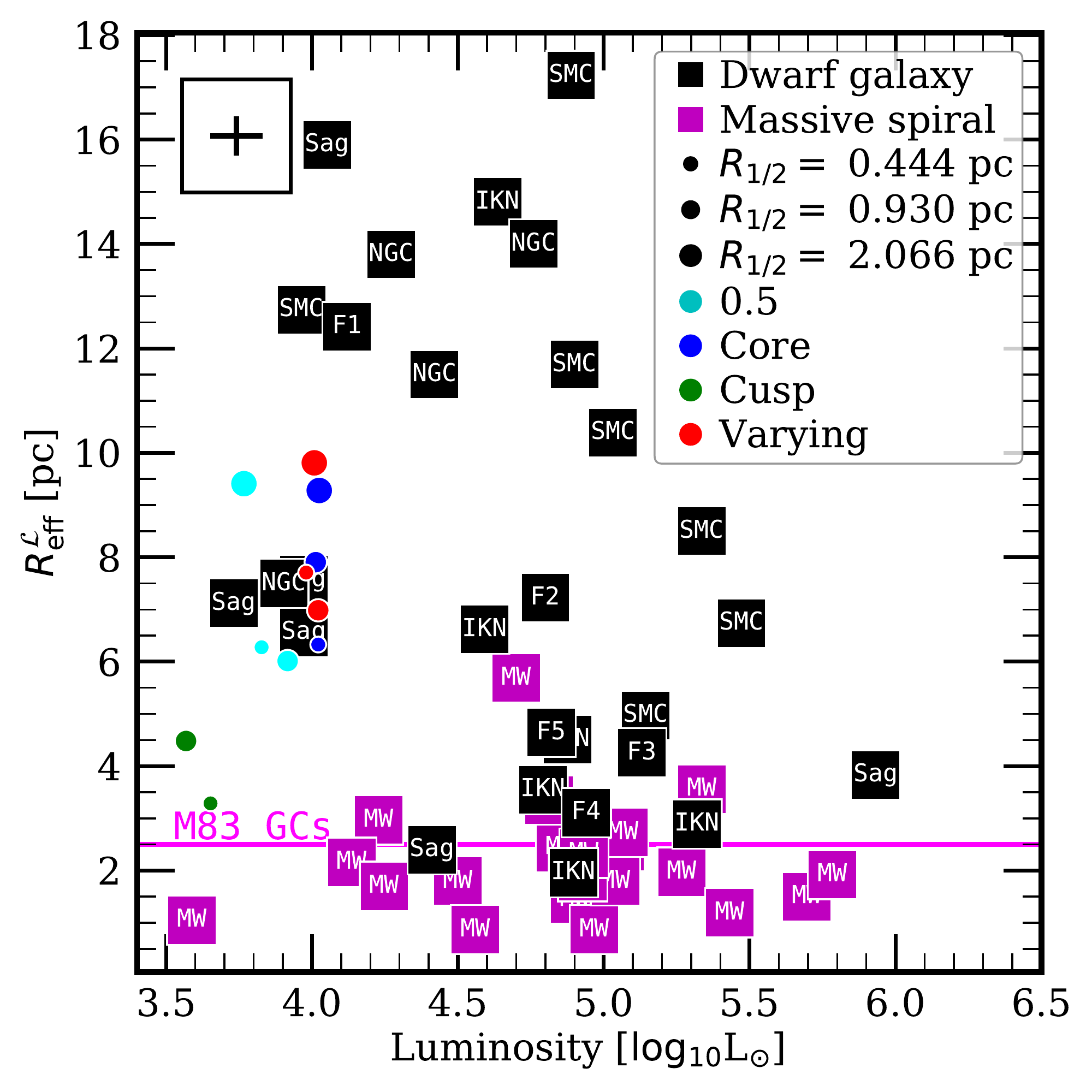} \\
  \end{tabularx}
\caption{Comparison of $R_{\rm eff}^\mathcal{L}$ and luminosity for the GCs simulated in this work and observed GCs in Fornax (F), NGC6822 (NGC), IKN (as marked), the Small Magellanic Cloud (SMC), Sagittarius (Sag), the Milky Way (MW; magenta data points) and M83 (magenta horizontal line). Massive spiral galaxies and dwarf galaxies are shown in magenta and black, respectively. Simulated values are taken at $\sim 14\,$Gyr. The top panel includes the 32k, 64k and 128k GC simulations at default densities; the bottom panel includes the 64k GC simulations at three different initial densities. Two of the 32k GC simulations and one of the 64k0.1 GC simulations are not displayed as they were fully destroyed prior to 14\,Gyr. A single error bar has been included in the top-left to indicate the average error bars of observed data.} \label{fig:fornax}
\end{figure}

A plot of the GC radius versus luminosity is shown in Fig. \ref{fig:fornax}. This plot shows a marked segregation between the GC sizes in dwarf galaxies and massive spiral galaxies. In our simulations, the GCs in $\gamma=0$ and $\gamma(t)$ potentials tended to have larger $R_{\rm eff}^\mathcal{L}$ (see Fig. \ref{fig:128krh} and Fig. \ref{fig:64krh}). The systematically larger size distribution of GCs in dwarf galaxies could therefore be indicative of a central dark matter core in these systems. Indeed, both from its stellar kinematics and the survival of its GC system, Fornax favours a large dark matter core \footnote{Note that the GC distribution of Fornax can be made compatible with a cuspy Fornax model if they originated far from their contemporary locations (\citealp{2012MNRAS.426..601C}; \citealp*{2019MNRAS.485.2546B}). However, in most models that achieve this, the GCs remain far from the centre of Fornax today which is unlikely given their distribution of projected distances \citep{2012MNRAS.426..601C}.} (\citealp{2006MNRAS.368.1073G, 2011MNRAS.411.2118A, 2011ApJ...742...20W, 2012MNRAS.426..601C, 2018MNRAS.480..927P, 2019MNRAS.482.5241K}; \citealp*{2019MNRAS.484.1401R}). Similarly, NGC 6822 favours a central core \citep*{2003MNRAS.340...12W}, though the analysis is complicated by the presence of a stellar and gaseous bar that makes the inner rotation curve asymmetric \citet{2016MNRAS.462.3628R}. \par

Note that the key parameter that determines the size-growth of GCs is their tidal field. The MW GCs that we consider here move in a stronger tidal field than those in the dwarfs and so are expected to have a smaller size. To quantify this, we calculate the radius in the MW that would have a similar tidal field to the GCs orbiting in our simulated dwarf. For this, we calculate the tidal radius for a point-mass (Equation \ref{tidal.eq} in the limit $r_{\rm s} \rightarrow 0$) with Galactocentric radius for the MW using the enclosed mass relationship for a generalised NFW profile (gNFW) presented in \citet*[Fig. 13]{2019MNRAS.485.3296W}. Based on this calculation, the tidal radii of all our simulated GCs shown in Table \ref{tab:sim1} would imply orbital radii of $6.63-19.80$\,kpc in the MW. To compare this to the orbits of the MW GCs in Fig. \ref{fig:fornax}, following \citet{2003MNRAS.340..227B} we define their mean orbital radius as $\langle R_{\rm g} \rangle \equiv R_{\rm pericentre}(1+e)$, where $e$ is the orbital eccentricity. Using values for the apocentre and pericentre of MW GCs in \citet[Table 1][]{2019MNRAS.482.5138B}, we find $\langle r \rangle = 2.62$\,kpc. This suggests that most MW GCs orbit in a stronger tidal field than our simulated GCs and, by extension, than GCs orbiting in real dwarf galaxies. However, we acknowledge that the orbital radius of a GC is also related to its mass \citep{2011MNRAS.413.2509G}, and most MW GCs are more massive than those we simulate here. \par

Finally, notice that there is a significant scatter in $R_{\rm eff}^\mathcal{L}$ of GCs in dwarf galaxies, varying from $\sim 2$\,pc to $\sim 18$\,pc (Note that the newly discovered GC in Sextans A is similarly large ($R_{\rm eff}^\mathcal{L} = 7.6 \pm 0.2$\,pc) \citep{2019MNRAS.487.1986B}.) This could owe to the GCs in dwarfs evolving to different sizes depending on their initial masses and orbits, or it could indicate that GCs in dwarfs are born with a wide distribution of sizes, unlike the young star clusters in M83 (magenta horizontal line in Fig. \ref{fig:fornax}) and the in-situ clusters in the Milky Way (magenta data points in Fig. \ref{fig:fornax}). Either way, the large GCs in dwarfs cannot orbit within a dark matter cusp as they would be rapidly destroyed \citep[e.g.][]{2012MNRAS.426..601C}. \par

% Comparison with Fornax GCs

\subsection{Can GCs retain memory of a historic cusp-core transformation?}

Several timescales are presented in \S\ref{sec:results} which play a role in whether GCs can `remember' their initial galactic potentials. These include the relaxation time $t_{\rm relax}$ (Equation \ref{relax.eq}), the cusp-core transformation time $t_{\rm tr}$ (\S\ref{sec:intro}) and the mass segregation time $t_{\rm seg}$ (discussed in \S\ref{sec:segre}). How these timescales relate to one another is an indication of whether the simulated GC properties will be degenerate after a Hubble time, or not. \par 

The most important timescale is the cusp-core transformation time, $t_{\rm tr}$. In the simulations we present here, the cusp-core transformation is complete by a Hubble time, and has reached $\gamma = 0.5$ by $\sim 8$\,Gyr. When compared with $t_{\rm seg}$ (which was found to be $\sim 1$\,Gyr) it can be seen that  mass segregation is complete long before the cusp-core transformation takes hold, and so the mass segregation is degenerate between the $\gamma(t)$ and static $\gamma$ simulations. \par

A similar comparison has been made with the relaxation time, using $\eta = \left \langle t_{\rm relax} \right \rangle /t_{\rm tr}$. If $\eta$ is too small then the GC will quickly adapt to its contemporary galactic potential profile and prior memory will be lost. Similarly, if $\eta$ is too large then the GC may not have had sufficient opportunity to relax into the initial galactic potential profile and will therefore lack any dynamical memory of it. \par

A promising result is the non-degeneracy of the GC $R_{\rm eff}^\mathcal{M}$ and $R_{\rm eff}^\mathcal{L}$ on the host galaxy potential slope. There are independent methods of measuring the potential slope in galaxies, such as analysis of galactic rotation curves (as demonstrated in (\citealt{1988ApJ...332L..33C}; \citealt*{2001AJ....122.2381M})). If a dwarf galaxy could be confidently identified as containing a DM core while hosting GCs with a lower than expected $R_{\rm eff}^\mathcal{L}$, then this would be evidence that the galaxy had experienced a historic cusp-core transformation. \par

Finally, the tidal debris stripped from the GCs constitutes the most reliable test of an historic cusp core transformation (see Fig. \ref{fig:vis}). Due to the early presence of a cusp, stars are initially stripped from the GC, leading to more prominent tidal tails than can be produced in the static core simulations. The difference in mass in the tidal tails between the static core and time-varying simulations was large -- with the tidal tail mass of the time-varying simulation greater by nearly a factor 2. This suggests that this probe will prove effective in practice. A caveat, however, is that these tests are only meaningful when observers are able to compare the tidal debris and size of GCs with predictions for these properties in static host potentials. Otherwise, an observer would be unable to claim that a GC had a smaller than expected $R_{\rm eff}^\mathcal{L}$ or greater than expected tidal debris. This is further complicated when one considers that the rate of mass-loss in static cusps is sensitive to the steepness of the potential slope \citep*{2017MNRAS.466.3937C}, and that the retention fraction of BHs affects both the GC mass-loss rate and size evolution (\citealp{2007MNRAS.379L..40M, 2016MNRAS.462.2333P, 2017MNRAS.470.2736P}; \citealp*{2018MNRAS.479.4652A}). Loose predictions could be made with $N$-body simulations, or with tools such as {\sc emacss} \citep{2014MNRAS.442.1265A}, but these estimates would be subject to their own uncertainties and one would need to know the mass distribution of the host galaxy. \par

\section{Conclusions} \label{sec:conclusion}

We have used the {\sc nbody6df} code to simulate the dynamical evolution of GCs orbiting in dwarf galaxies, with the goal of addressing the question: do GCs retain a dynamical memory of dark matter cusp-core transformations? Our simulations included the effect of stellar evolution, modelling GCs of different initial mass and density orbiting in a static dark matter cusp ($\gamma=1$), a static dark matter core ($\gamma=0$) and a dark matter halo that slowly transformed from being cusped to cored over a Hubble time ($\gamma(t)$). We also considered the effect of different initial GC particle number and density. Our key findings are as follows: \par

\begin{enumerate}[label=(\roman*),leftmargin=*,align=left,labelsep=0mm]

\item The evolution of each GCs' $R_{\rm eff}^\mathcal{M}$ (as shown in Fig. \ref{fig:128krh} and Fig. \ref{fig:64krh}) shows that GC size is dependent on $\gamma$ -- the slope of a host galaxy's potential profile. \par

\item The size of a GC after a Hubble time is distinct in each of our simulations. We found that GCs orbiting in a static dark matter cusp were substantially smaller (by $\sim 2-4$\,pc) than those orbiting in the static core or time-varying potentials ($\gamma(t)$). Furthermore, the GC orbiting in the $\gamma(t)$ potential had a smaller $R_{\rm eff}^\mathcal{M}$ than that of the GC in a static core. However, this difference would be challenging to unpick in practice given the unknown initial size, mass and orbit of the GC.

\item The different dwarf galaxy mass profiles considered here were not distinguished by the mass segregation of their GCs or the velocity dispersion within the GC $R_{\rm eff}^\mathcal{M}$. All simulated GCs were maximally mass segregated within $\sim 1$\,Gyr, by which time the dynamical properties of the GC did not have sufficient time to respond to their host galactic potential. \par

\item The cleanest signature of an historic cusp-core transformation was found to be the presence of large GCs surrounded by tidal debris. It remains to be seen, however, if this signature can be unambiguously extracted from real data, given the uncertain orbit, initial size and initial mass of the GC. We will consider this in future work.\par

\item Finally, we compared our simulated GCs with observed GCs in nearby dwarf galaxies (Fig. \ref{fig:fornax}). We found that GCs in dwarf galaxies are larger (by an average of $\sim 5$\,pc) and exhibit greater size scatter than those formed in-situ to the Milky Way and M83. Such large GCs form and survive naturally in our static dark matter core simulations and in our simulations in which a dark matter cusp is slowly transformed to a core over a Hubble time, but not in our simulations of dwarf galaxies with a dark matter cusp. This suggests that, while nearby dwarf galaxies may have had a dark matter cusp in the past, they do not have a dark matter cusp today. \par

\end{enumerate}

\section{Acknowledgements}

Mark Gieles acknowledges financial support from the European Research Council (ERC-StG-335936, CLUSTERS). We are grateful to Sverre Aarseth and Keigo Nitadori for making {\sc nbody6} publicly available. We also thank Mr. Dave Munro of the University of Surrey for hardware and software support.
This research has made use of the NASA/IPAC Extragalactic Database (NED) which is operated by the Jet Propulsion Laboratory, California Institute of Technology, under contract with the National Aeronautics and Space Administration.
Finally, we would like to thank the referee, Douglas Heggie, for his constructive and helpful review that improved the clarity of this work.

\bibliographystyle{mnras}

\setlength{\bibhang}{10pt}
\bibliography{references} \label{Bibliography}

\end{document}